\documentclass[fleqn,10pt]{wlscirep} 
\usepackage[utf8]{inputenc} 
\usepackage[T1]{fontenc} 

\usepackage{color,soul}

\usepackage{amsmath,amssymb} 
\usepackage{changepage} 
\usepackage{textcomp,marvosym} 
\usepackage{cite} 
\usepackage{nameref} 
\usepackage[right]{lineno} 
\usepackage{appendix,braket,booktabs,array} 

\title{Long-term tracking of social structure in groups of rats}

\author[1,2,3,4,5,6,$\dagger$*]{Máté Nagy}
\author[4,5,6,$\dagger$*]{Jacob D. Davidson}
\author[1,3]{Gábor Vásárhelyi}
\author[1]{Dániel Ábel}
\author[7,8,9]{Enikő Kubinyi}
\author[4,5,6]{Ahmed El Hady}
\author[1,3]{Tamás Vicsek}
\affil[1]{Department of Biological Physics, Eötvös Loránd University, Budapest, Hungary}
\affil[2]{MTA-ELTE ‘Lendület’ Collective Behaviour Research Group, Hungarian Academy of Sciences, Budapest, Hungary}
\affil[3]{MTA-ELTE Statistical and Biological Physics Research Group, Hungarian Academy of Sciences, Budapest, Hungary}
\affil[4]{Department of Collective Behaviour, Max Planck Institute of Animal Behavior, Konstanz, Germany}
\affil[5]{Department of Biology, University of Konstanz, Konstanz, Germany}
\affil[6]{Centre for the Advanced Study of Collective Behaviour, University of Konstanz, Konstanz, Germany}
\affil[7]{Department of Ethology, Eötvös Loránd University, Budapest, Hungary}
\affil[8]{ELTE NAP Canine Brain Research Group, Budapest, Hungary}
\affil[9]{MTA-ELTE Lendület ‘Momentum’ Companion Animal Research Group, Budapest, Hungary}

\affil[$\dagger$]{These authors contributed equally to this work}
\affil[*]{nagymate@hal.elte.hu, jdavidson@ab.mpg.de}

\usepackage{setspace} 

\begin{abstract}
Rodents serve as an important model for examining both individual and collective behavior. Dominance within rodent social structures can determine access to critical resources, such as food and mating opportunities. Yet, many aspects of the intricate interplay between individual behaviors and the resulting group social hierarchy, especially its evolution over time, remain unexplored. In this study, we utilized an automated tracking system that continuously monitored groups of male rats for over 250 days to enable an in-depth analysis of individual behavior and the overarching group dynamic. We describe the evolution of social structures within a group and additionally investigate how past behaviors influence the emergence of new social hierarchies when group composition and experimental area changes. Notably, we find that conventional individual and pairwise tests exhibit a weak correlation with group behavior, highlighting their limited accuracy in predicting behavioral outcomes in a collective context. These results emphasize the context-dependence of social behavior as an emergent property of interactions within a group and highlight the need to measure and quantify social behavior in more naturalistic environments.
\end{abstract}
\begin{document}

\flushbottom
\maketitle


\section*{Introduction}
Collective behavior emerges based on interactions between individuals in a group.  This is observed at many different scales, from wound healing at the cellular level \cite{vishwakarma_mechanical_2018}, to task allocation in social insects \cite{davidson_effect_2016}, group search behavior \cite{nagy_synergistic_2020}, and information exchange on human social networks \cite{bond_61-million-person_2012}.  
Hierarchical structures are common in animal groups, for example, in the grooming relationships of chimpanzees \cite{kaburu_egalitarian_2015},
leadership and movement of pigeons \cite{nagy_context-dependent_2013}, and reproduction in cichlid fish \cite{maruska_social_2014}.
Social network structure influences decision-making \cite{king_dominance_2008,stewart_information_2019}, and dominance position within a network can influence an individual's fitness \cite{majolo_fitness-related_2012}.

With rats, previous work has shown that individuals within a group have a social status related to dominance and that aggression and avoidance behavior are key elements of social interactions \cite{barnett_analysis_1958,grant_analysis_1963,schweinfurth_social_2020}.  
However, it is not known how individual interactions lead to the overall social structure of the group and how social structures change over time \cite{dennis_systems_2021}.
Fortunately, new automated tracking methods enable long-term tracking of individuals within social groups, providing a quantitative description of behavior and interactions \cite{forkosh_animal_2021}.
For example, recent work has analyzed the ontogeny of collective behavior in zebrafish \cite{hinz_ontogeny_2017}, lifetime behavioral differences in honeybees \cite{smith_behavioral_2022}, and how genetic relatedness corresponds to group social structure in mice \cite{evans_long-term_2021}. 

While it is acknowledged that long-term multi-modal characterizations are required to describe complex social behaviors, there are still a number of challenges \cite{jabarin_beyond_2022}.
An approach applicable to both lab and field conditions is to tabulate appropriate pairwise interaction information of animals in groups and use this to define measures of an individual's position in the social hierarchy.
With mice, both aggressive and non-aggressive interactions have been used to define dominance hierarchies
\cite{shemesh_high-order_2013,forkosh_identity_2019,karamihalev_social_2020,lopez_ketamine_2022,williamson_temporal_2016,lee_effect_2021,lee_distinct_2022}.
With primates, for example, pairwise ``supplanting'' interactions, such as when one individual displaces another from a food source, have been used to determine an individual's ranking \cite{johnson_supplanting_1989,evans_inferring_2018,gullstrand_computerized_2021}.  

Automated identification of approach-avoidance interactions has been used in previous work as a scalable method to characterize group hierarchy and leader-follower relationships \cite{nagy_context-dependent_2013}.
Previous work with rodents, including rats \cite{spruijt_approach_1992} and mice \cite{forkosh_animal_2021}, has also considered approach and avoidance events.
An approach-avoidance event occurs when one individual approaches another, but the other individual moves away (either by retreating or escaping).

There are multiple ways to compute social rankings based on pairwise interactions such as approach-avoidance events. The Elo score, which was originally developed to rank chess players and predict the outcome of future matches \cite{elo_rating_1978}, is commonly used in animal behavior to calculate an individual's ranking using pairwise contest or interaction information \cite{albers_elo-rating_2001,neumann_assessing_2011,strauss_inferring_2019}.
If network ``flow'' is defined from winners to losers in an interaction network, the network measure of ``flow hierarchy,'' which refers to how information flows through the network, can also be used to quantify hierarchical structure.
The local reaching centrality (LRC) considers flow hierarchy in a network and quantifies a node's ability to efficiently reach other nodes in its immediate neighborhood \cite{mones_hierarchy_2012}.
Since ``flow'' occurs in the direction from winner to loser (or dominant to less-dominant), dominant individuals have higher local reaching centrality.
Global reaching centrality (GRC) is calculated using the distribution of LRC scores, and this metric has been used to quantify the steepness of hierarchical structures in many different systems, including groups of horses \cite{ozogany_modeling_2015}, ant colonies \cite{shimoji_global_2014}, brain networks \cite{kora_global_2023}, industrial trade networks \cite{hu_hierarchy_2017,beardsley_hierarchy_2020}, and scientific citation networks \cite{mones_universal_2014}. 
In animal behavior, in addition to the individual Elo scores, metrics employed to measure social dominance hierarchies also include directional consistency of contests or interactions, proportions of interactions based on rank, transitivity, linearity of the hierarchy, and David's score or Elo score distributions across the group \cite{strauss_domarchive_2022,williamson_temporal_2016,shizuka_social_2012,de_vries_improved_1995,neumann_extending_2023}.

In this work, we developed an open-source vision-based automated tracking and behavioral characterization system to analyze the social behavior of small animals like rodents.
We use this system to continuously monitor interactions and behavior of rat groups, enabling us to quantitatively examine the temporal evolution of social structure and the roles of individuals within these groups.
We note, however, that rodent groups have inherently complex behavior and social structures, making it difficult to draw broad conclusions about the rules and processes that govern observed structures. 
We, therefore, focus on characterizing the results of our experiments and on using multiple metrics to describe various dimensions of the social structure.

For an extended 36-week period, we tracked the social behavior of 28 rats divided into several groups, and calculated behavioral metrics and interactions to analyze both individual and group behavior. 
We examined how individual behavioral differences persist and combine to form new social structures when the composition of the group is altered.
At first, the rats were divided into 4 groups of 7, and following this, we merged groups. For the final sequence of group experiments, we then created 4 new groups of 7 and altered the size of the living areas.
Following the completion of the group experiments, we ran individual behavioral assays on each rat and compared the results to those of the group experiments.
This combination of methods and experiments enabled us to (1) identify a wide range of individual locomotion and social behaviors, (2) investigate the formation and details of dominance hierarchies, (3) investigate the effect of group composition changes and the associated social stress on the behavior of rats living in groups in enriched environments, and (4) compare behavioral assay results to behavior observed in a group setting.
Overall, our work demonstrates scalable methods for describing long-term changes in animal group social structure and emphasizes the need to use such methods to obtain a full picture of group social structure and interactions in natural or semi-natural environments.

\section*{Results}

\subsection*{Long-term tracking and quantifying individual behavior}
\begin{figure}
    \centering
    \includegraphics[width=\linewidth]{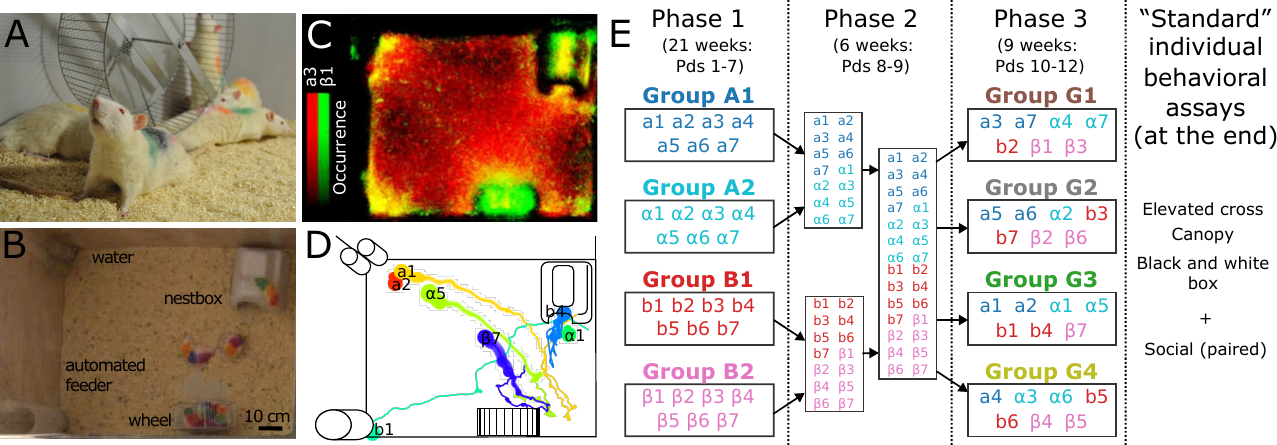}  
    \caption{\textbf{Experiment setup and timeline.}
(A) Photo of the rats with color-codes for individual identification and tracking. 
(B) Still image from the video that was used for tracking (from group $G1$, during Pd 10) taken by a light-sensitive camera at low lighting conditions. Image overlaid with labels indicating the important objects (water, nestbox, etc.).
(C) Continuous tracking allowed for the reconstruction of each individual's space use. The heatmap shows the space use of two rats during a 3-week period at the beginning of phase 3. Areas used only by $a3$ are shown with red, only by $\beta1$ with green, and areas visited by both (e.g. at the water and the feeder) are shown with yellow.
(D) Trajectories were used to identify dominance interactions in the form of approach-avoidance events, where one individual approaches another, but the other moves away (by backing up or fleeing).  Shown is an example of trajectories from group $G3$ in period 10.  
Lines show locations for 60 seconds, with the semitransparent circles of increasing size showing the more recent positions.
(E) Overview of experimental manipulations.  We calculate behavioral metrics over each 3-week "period" (abbreviated as Pd).  Phase 1 had rats in original breeding line-sorted groups $A1$-$A2$ (line $A$), $B1$-$B2$ (line $B$), for a total of 7 periods.
Each rat is labeled with lowercase letters $a$/$\alpha$ or $b$/$\beta$ according to breeding line. Individual numbers within each group are sorted in ascending order according to rank as determined by Elo score at the end of phase 1, i.e.\ $a1$/$a7$ were the highest/lowest ranking individuals in $A1$ during Pd 7, $\alpha1$/$\alpha7$ were the highest/lowest in $A2$, etc.
In phase 2, the groups were mixed together by breeding line during Pd 8, and then all together for Pd 9.  At the beginning of phase 3 (Pd 10), new groups were formed ($G1$-4).  During Pds 11 and 12 in phase 3, the compartment area sizes were changed (see Methods and Fig \ref{sfig:arenasize}). At the end of the experiments, individual behavior was assessed by traditional individual and pairwise assays.
}
    \label{fig:illustrate}
\end{figure}

We tagged individuals with color markers and employed automated tracking to determine each rat's movement over time (Figure \ref{fig:illustrate}A-D).
Over the course of the experiment, we performed manipulations to alter the group composition and the living area available for the group to use. 
We used two different {breeding lines of Wistar} laboratory rats, denoted $A$ and $B$ (see Methods for details), with associated individual labels $a*$ or $\alpha*$ and $b*$ or $\beta*$, respectively (see Figure \ref{fig:illustrate}E).
We initially divided the rats into four groups of seven, with $A$ rats in groups $A1$ and $A2$ and $B$ rats in groups $B1$ and $B2$.
The rats remained in these groups for the first observation period, which lasted a total of 21 weeks - we denote this time as phase 1.
Following this, in phase 2, we merged groups $A1$-$A2$ and $B1$-$B2$ for three weeks and then merged all for three weeks by opening portals between their compartments.
For the final series of group experiments in phase 3,
rats from each original group were mixed together to create four new groups.
The reshuffling in phase 3 was done according to body mass at the end of phase 2 { (mean = 480 g; min = 364 g, max = 613 g; Q1 = 423 g, median = 481 g, Q3 = 532 g)} , allocating rats to new groups by ensuring that each group had the full range of masses and included members from every previous group {($G1_{min}$ = 394 g,
$G1_{max}$ = 541 g,
$G2_{min}$ = 372 g,
$G2_{max}$ = 587 g,
$G3_{min}$ = 400 g,
$G3_{max}$ = 613 g,
$G4_{min}$ = 364 g,
$G4_{max}$ = 555 g)}.
Figure \ref{fig:illustrate}E shows the experimental structure and the associated measurement periods.

We calculated automated trajectory-based behavioral metrics to quantify behavior over the duration of the experiment.
We calculated and averaged each metric over successive time periods of 3 weeks (denoted as Pd), with associated numbers 1-12. 
We use the summary metrics to ask how behavior changes over time, how individuals differ, how groups differ, and how previous individual and group behavior predicts changes when new groups are formed.

To assess dominance-related interactions and social structure, we tabulated approach-avoidance events between all pairs of rodents in each group. This automated method defines ``events'' as when a pair of rats come close to each other: the ``displacer'', i.e.\ the dominant rat in an event, subsequently stays in place or continues moving forward, while the other (the ``displaced'', i.e.\ subordinate rat) move away \cite{nagy_context-dependent_2013}.
This type of approach-avoidance interaction can also be dynamic, such as when one individual chases another.
We use the matrix of approach-avoidance events to calculate metrics that describe the dominance structure of each group and each individual's position in this structure.

\subsection*{Breeding line and group differences}
We use automated measures of space use and pairwise interaction events to characterize individual and group behavior. We first examine general differences between breeding lines.

In the beginning, the rats were juveniles and were growing rapidly, as shown by the large increases in body mass during this time period. 
The $A$ rats were, on average, significantly larger than the $B$ rats during each period (T-test comparing average mass of $A$ rats to $B$ rats yields $p<0.001$ for each period).
All rats had approach-avoidance events during the experiment, and there were no consistent significant differences among the breeding lines.
However, there was an increase in the number of events per rat in phase 3 compared to phases 1 and 2 (Mean number of events per rat in phases 1, 2, 3, respectively: 447, 494, 976; T-test mean of phase 1 to phase 2, p=0.64; mean of phase 1 to phase 3, p=0.0044; mean of phase 2 to phase 3: p=0.0176).

\begin{figure}
    \centering
     \includegraphics[width=0.85\linewidth]{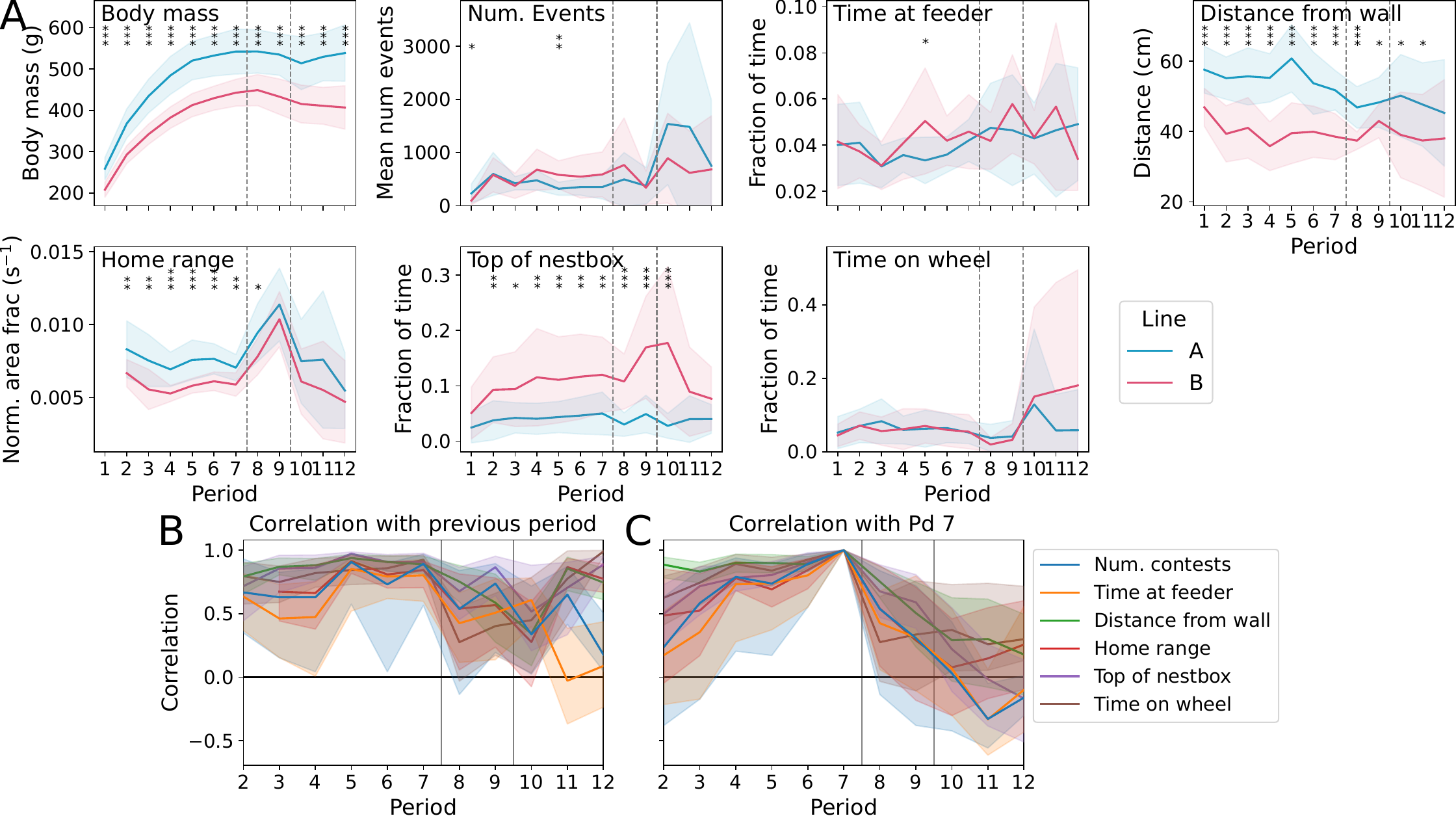}
    \caption{\textbf{Breeding line comparison and correlation}.
    (A) Per-line body mass, average number of events, and space use metrics.  Significant differences between breeding lines for a designated period, as determined with a T-test for difference in means, are denoted as follows: $p<0.05$ with *, $p<0.01$ with ** and $p<0.001$ with ***.
    See also Fig \ref{sfig:groupmetrics} for space use compared according to group, and Fig \ref{sfig:indivspaceuse}) for space use metrics for each individual rat.
    (B) Correlation with the previous period, calculated across all rats with respect to a particular metric. Shaded area shows confidence interval calculated via bootstrapping.  Note that values significantly different from zero are when the confidence intervals do not contain zero.
    (C) Correlation with Pd 7 (the last measurement in phase 1). Shaded area shows confidence interval calculated via bootstrapping.
    }
    \label{fig:linecomp_and_corr}
\end{figure}

The metrics of time at feeder, distance from wall, home range\cite{burt1943territoriality}, time at top of nestbox, and time on wheel describe space use.  While the breeding lines did not have general differences in time at feeder or time on wheel, line $A$ rats tended to be farther from the wall, visited more parts of the living compartment (larger home range),  and spent less time on top of the nestbox in comparison with $B$ rat groups.  However, while these differences were clear during phase 1, the differences in distance from wall and home range decreased when the lines were mixed, with home range no longer significantly different from Pd 9 onward, and distance from wall no longer significant in Pd 12.  Breeding line differences in time spent at top of nestbox showed a large increase when group membership was changed in Pds 9 and 10, but subsequently decreased and were not significant in Pds 11 and 12 (Fig \ref{fig:linecomp_and_corr}). 
Note that one group (G1) in phase 3 displayed a different pattern of wheel usage than other groups, with several rats spending a very large amount of time on the wheel at the same time and thus unable to use it for running (Figs \ref{sfig:groupmetrics} and \ref{sfig:indivspaceuse}); however, there were no breeding line differences in this behavior.
Overall these metrics suggest that the different breeding lines differed in their space use tendencies, but differences decreased when rats were placed in mixed groups in phase 3.

We quantify changes in individual behavior using the correlation coefficient across periods.
This shows that individuals have consistency in number of events and space use, as demonstrated by the generally positive correlations during the entire observation period (Fig \ref{fig:linecomp_and_corr}B).  
However, while there is consistency from one period to the next, Fig \ref{fig:linecomp_and_corr}C shows that small behavioral shifts over time can accumulate.  Moreover, we see that the re-groupings facilitated changes in behavior. This is demonstrated by the sharper decrease in the correlation of behavioral metrics with Pd 7 in phases 2 and 3 compared to that in phase 1 preceding Pd 7.
In particular, while the correlation coefficient for home range and time at top of nestbox during Pds 11 and 12 showed high correlations (Fig \ref{fig:linecomp_and_corr}A), the correlation of these 3 measurements with Pd 7 values was lower (Fig \ref{fig:linecomp_and_corr}B).
For example, for Pd 12, the correlation with the previous period for home range was 0.77 (95\% CI: [0.68 0.87] and for top of nestbox was 0.89 (95\% CI: [0.71 0.95]), while the correlation values with Pd 7 were 0.26 (95\% CI: [-0.13  0.57]) and -0.17 (95\% CI: [-0.5   0.16]), respectively.
This indicates that the new behavioral routines of phase 3 differed from those of phase 1.

\subsection*{Metrics for group social structures}
\begin{figure}
    \centering

    \newcolumntype{P}[1]{>{\centering\arraybackslash}m{#1}}
    \begin{tabular}{cP{1.6cm}P{1.6cm}P{1.6cm}P{1.6cm}P{1.6cm}P{1.6cm}P{1.6cm}P{1.6cm}}
    \toprule
    &   Connected hierarchy &Line  &Layered hierarchy & Layered-half & Non-transitive & Single dominant & Single out & Symmetric \\
    \midrule 
    ESS   &   \textbf{1.00} &\textbf{1.00} &0.81 & 0.79 & 0.65 & 0.59 & 0.59 & \textit{0.55} \\
    DSS   &   \textbf{0.99} &0.08 &0.41 & 0.32 & 0.12 & 0.11 & 0.11 & \textit{0.00} \\
    GRC &   0.58 &0.58 &0.78 & \textbf{1.00} & 0.00 & \textbf{1.00} & \textit{0.03} & - \\    
    DCI   &   \textbf{1.00} &\textbf{1.00} &\textbf{1.00} & \textbf{1.00} & \textbf{1.00} & \textbf{1.00} & \textbf{1.00} & \textit{0.00} \\
    TTRI  &   \textbf{1.00} & - &\textbf{1.00} & \textbf{1.00} & \textit{0.00} & - & - & - \\
    \bottomrule
    \end{tabular}    
    \vspace{6pt}\\
    \includegraphics[width=\linewidth]{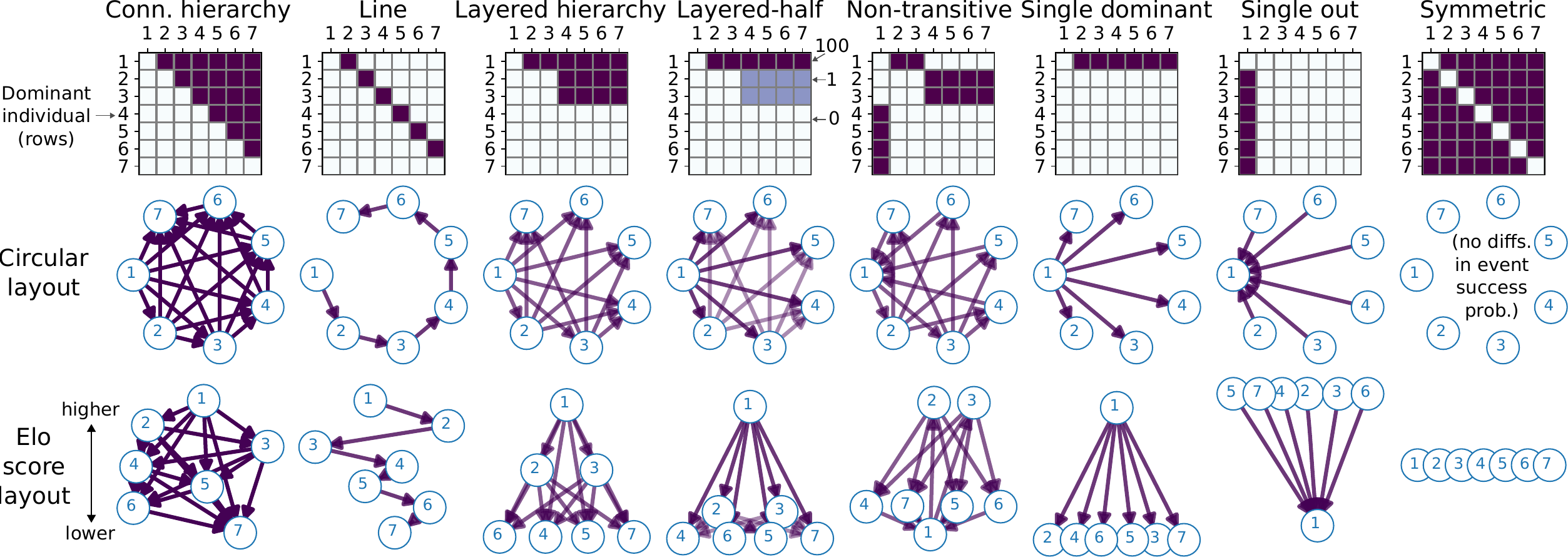}
    \caption{\textbf{Idealized networks and group social metrics}. 
    The table at the top shows the scores calculated: Elo score steepness (ESS), David's score steepness (DSS), global reaching centrality (GRC), directional consistency index (DCI), and triangle transitivity index (TTRI).  Note that the GRC is not defined for the symmetric network, and the TTRI is not defined for networks that do not contain any dominance triads.
    The different idealized networks have 7 nodes, and individual entries are either 100 or 0 (for the layered-half network, 100, 1, and 0 are used).
    The connected hierarchy network has a non-symmetric structure.
    The line network has a single ``line'' of pairwise interactions, where each individual only interacts with one other.
    The layered hierarchy network has a single individual who dominates all others and two other sub-dominant individuals who only dominate the four others below them. The layered-half network has the same structure but lower values for the subordinate individuals.  The non-transitive network has individual 1 dominating 2-3, 2-3 dominating 4-7, but 4-7 dominating 1.The single dominant network only has events with the dominant individual. The single-out network only has events with the subordinate individual.  The symmetric network has equal event dominance probability among all pairs. 
    In the table, the highest score for each metric (rows) is in bold, and the lowest score is in italics.
    The ESS is highest for the connected hierarchy and line networks, the DSS is highest for the connected hierarchy network while low for the line network, and the GRC is highest for the layered-half (i.e.\ structured but nonlinear hierarchy) and single dominant networks.
    For the symmetric network, the DCI is 0 because there are no consistent dominance relationships; for other networks, dominance is one-sided and the DCI is 1.
    The TTRI is 0 for the non-transitive network, and 1 for other networks where dominance triads are predicted.
    In addition to matrix plots, each network is visualized by showing connections in the direction of the more to less dominant individuals in each pair (note that no connections are shown for the symmetric network because, in this case, there are no differences in event dominance probability). We show both a circular layout and a layout based on Elo scores, where individual nodes with higher Elo scores are shown higher up on the $y$-axis.
    }
    \label{fig:idealnetworks}
\end{figure}

With the pairwise approach-avoidance interaction matrices for each period, we use multiple metrics to characterize different aspects of group social structure and an individual's placement in this structure. The metrics to characterize individual social placement include Elo score, David's score,  local reaching centrality, and fraction of events dominated, and those to characterize group social structure include Elo score steepness, David's score steepness, global reaching centrality, directional consistency index, and triangle transitivity index. In this section we use idealized networks (shown in Fig \ref{fig:idealnetworks}) to illustrate what the group social structure metrics represent.  Note that while other work has used similar idealized or artificial networks as ``categories'' to label group social structure \cite{varholick_social_2019}, here we use the ideal networks (including connected hierarchy, line, layered hierarchy, layered-half, non-transitive, single dominant, single out, and symmetric) not as categories, but rather to give intuition for how the different metrics describe different aspects of the social structure.  In the following section, we report the metrics for each group and use them to describe the experimentally observed structures.

The Elo score steepness (ESS) is a measure of the spread of the distribution of Elo scores across the group.
It is calculated by converting the Elo score to a success probability, summing normalized values across group members, and calculating the slope of a linear regression fit to the resulting values \cite{neumann_extending_2023}.
The David's score steepness (DSS) (often referred to simply as hierarchy `steepness', or `classic steepness' \cite{strauss_domarchive_2022,neumann_extending_2023}) is calculated as the slope of a linear regression fit to the normalized David's scores among group members \cite{de_vries_measuring_2006}.
Individual local reaching centrality (LRC) uses the directed network of excess pairwise event outcomes (positive entries for rats in a pair that was dominant in more events, and zero for the other rat - see Methods) in order to assign higher scores to individuals in higher positions within a group hierarchy.
For an unweighted directed network, LRC is the fraction of nodes reachable by any given node; a generalization of the metric accounts for weighted connections \cite{mones_hierarchy_2012}. 
Global reaching centrality (GRC) is the average difference of nodal LRC with that of the highest LRC of any node in the graph, and a higher GRC indicates a more hierarchical network \cite{mones_hierarchy_2012}.

The directional consistency index (DCI) is the fraction of events dominated by the more dominant individual of each pair, with 1 corresponding to perfect predictability in the outcome of a pairwise event (i.e.\ one individual is always dominant), and 0 representing an exchange of approach-avoidance outcomes (i.e.\ each individual dominates the same number of events) \cite{van_hooff_dominance_1987,strauss_domarchive_2022}.
The triangle transitivity index (TTRI) is the fraction of triad relationships that show transitivity in pairwise event dominance outcomes (i.e.\ if $a\rightarrow b$ and $b \rightarrow c$, then $a \rightarrow c$ for a transitive triad)  \cite{shizuka_social_2012,shizuka_network_2015}.  

From Fig \ref{fig:idealnetworks} we note that the ESS and DSS, which both aim to measure the steepness of hierarchy within a group, show similar trends at times and differ at others; both have high values for the connected hierarchy network but differ for the line network.  
We also note that the aspects of the network structure described by ESS/DSS versus GRC are different (c.f.\ differences in the connected hierarchy, layered-half, and single dominant networks); the former is maximized when a well-connected structure exists (i.e.\ the hierarchy shows a clear distribution that lends itself to a linear regression fit), while the latter is maximized when more extremes in hierarchical structures exist (for example, the single dominant).
Although a comprehensive evaluation of these metrics is beyond the scope of this study (see, for example, \cite{neumann_extending_2023}), here we calculate and examine multiple metrics to ensure a robust interpretation of the data, as well as to facilitate comparison of our findings with other assessments of group social structure found in the literature.

\subsection*{Group social structures}

\begin{figure}
	\centering
	\includegraphics[width=\linewidth]{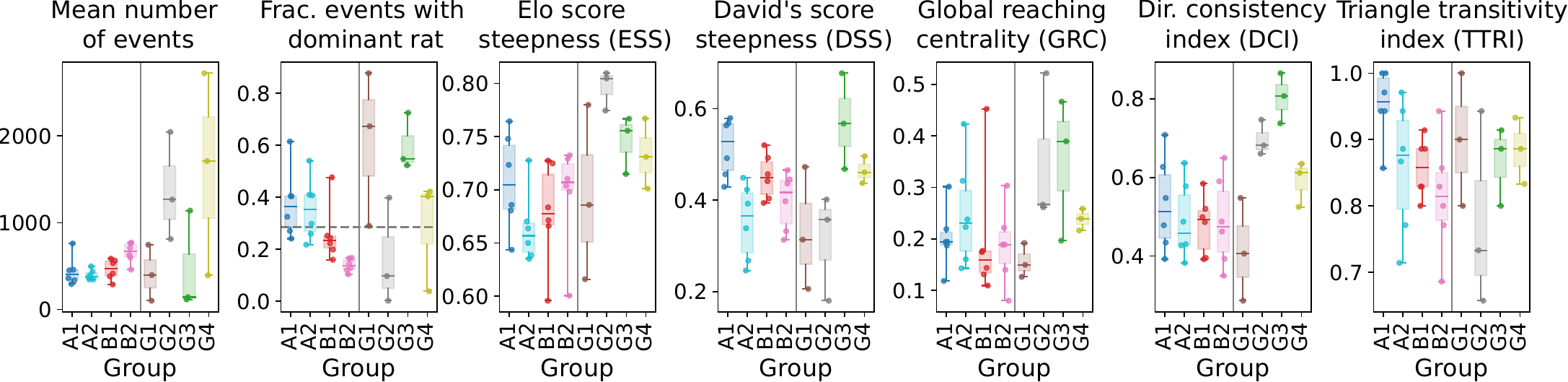}
	\caption{\textbf{Measures of group social structure}.
    The mean number of pairwise events and the fraction of total events with the most dominant rat are shown in addition to the hierarchy-related metrics of Elo score steepness (ESS), David's score steepness (DSS), global reaching centrality (GRC), directional consistency index (DCI), and triangle transitivity index (TTRI). The fraction of events with the dominant rat (where ``dominant'' is defined as the individual with the highest Elo score) is analogous to the measure of ``despotism'' used in other work \cite{karamihalev_social_2020}; the dashed line shows the expected value if all pairs of rats have the same number of events.
    The metrics are calculated for each period and are shown as boxplots for each phase 1 and phase 3 group.  See also Fig \ref{sfig:gnetworkovertime} for values for each group over time.
	}
	\label{fig:groupnetwork}
\end{figure}

\begin{figure}
    \centering
     \includegraphics[width=\linewidth]{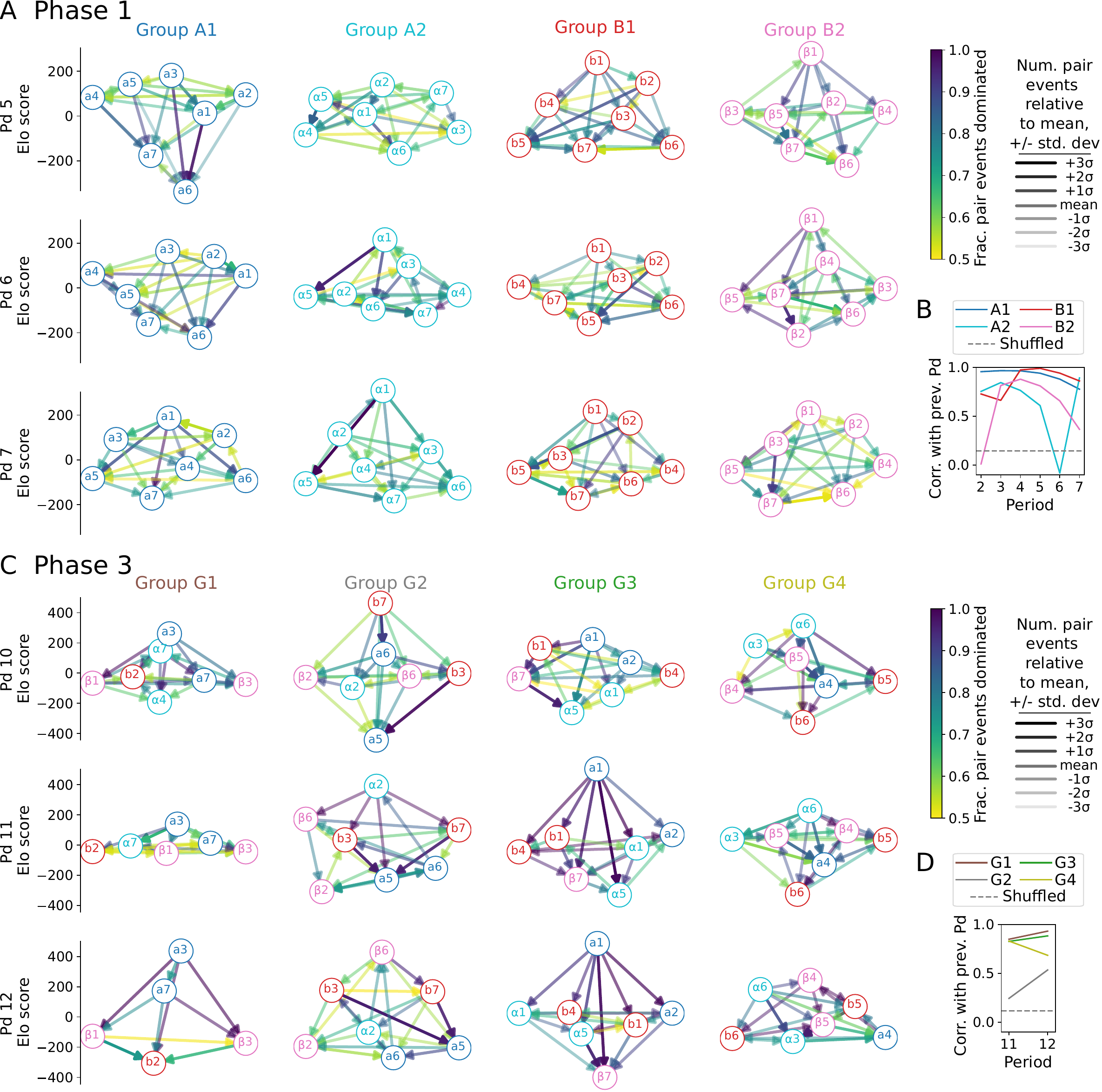}
    \caption{\textbf{Social structure network visualizations}.
    (A,C) A visualization of group networks during (A) the last 3 periods of phase 1, and (C) phase 3.
    Columns correspond to different groups and rows for each period. The position of each individual on the $y$-axis is set according to their Elo score.
	The direction of each connection indicates which individual dominated more events in the pair (e.g.\ a connection $a4\rightarrow a6$ indicates that $a4$ more often displaced $a6$ than vice versa), the color indicates the fraction of events dominated, and the transparency is proportional to the total number of pairwise events relative to the mean for that group and period.
    (B,D) The correlation of individual Elo scores with the previous period, for (B) phase 1 groups, and (D) phase 3 groups. Dashed line shows baseline correlation value calculated by shuffling groups and periods during phase 1 (B) or phase 3 (D).
    See also Fig \ref{sfig:indivnetworkmetrics} for individual metrics (including num.\ events, fraction dominated, Elo score, David's score, and reaching centrality) for each individual rat plotted for each period.
    }
    \label{fig:networks}
\end{figure}

We find that groups differ in their social structure, but within-group structure shows consistency when group membership remains unchanged.
We compare phase 1 and phase 3 group social structures because the associated periods all featured groups of 7 rats. We show results for all group social structure metrics, as well as the mean number of events and fraction of events with the dominant rat, and note instances where the trends for the metrics are similar versus contrasting.

In phase 1, in contrast to space use, which showed clear breeding line-based differences (Fig \ref{fig:linecomp_and_corr}), we do not see clear differences between lines $A$ and $B$ in terms of overall group social structure.  While the fraction of events with the dominant rat was higher for the $A$ groups in comparison to the $B$ groups, other metrics do not show large or consistent differences (Fig \ref{fig:groupnetwork}).
Within phase 1, each group showed consistency in the social structure over time, with the exception of Pd 6 for group $A2$, where a large change in the individual rank ordering in the social structure of the group took place. This is seen in the network visualizations, as well as in the positive correlation of Elo scores from one period to the next (Fig \ref{fig:networks}A-B).

In general, we saw larger differences between the groups in phase 3 in comparison to those in phase 1.
In phase 3, $G1$ and $G3$ each had a single consistent dominant individual, $G2$ had ongoing changes in social structure, and $G4$ had a stable hierarchy but with ongoing events.
All groups had consistency in structure, but the correlation of individual scores with the groups was higher for groups $G1$, $G3$, and $G4$ in comparison with group $G2$.  (Fig \ref{fig:networks}C-D).

Compared to other phase 3 groups, $G1$ and $G3$ had a relatively low number of events and a high fraction of events with the dominant individual.
Each of these groups had a single individual that was consistently ranked as most dominant (see Fig \ref{fig:networks}C).
However, in comparison to $G1$, $G3$ had on average higher DSS, ESS, and GRC.  This and the higher DCI index suggests that group $G3$ had a steeper hierarchical structure than group $G1$.

In contrast to groups $G1$ and $G3$, group $G2$ did not have a single individual that remained dominant during each period.  Group $G2$ had many events, the lowest fraction of events with the dominant rat, and the lowest transitivity (TTRI) of the groups. This and the lower correlation coefficient in Elo scores compared to other phase 3 groups suggest an ongoing struggle for position within the social network where ongoing events maintained pairwise relationships.
However, we note the differences obtained in the hierarchy steepness measures for group $G2$: the David's score steepness suggests a weak hierarchy, while the Elo score steepness and GRC suggest hierarchies definitely exist. 

Group $G4$ had similar patterns of metrics to Group $G3$, but with several distinct differences: these include lower magnitudes of ESS, DSS, GRC, and DCI, a lower fraction of events with the dominant individual, and overall many more events (although the mean number of events decreased dramatically from Pd 11 to Pd 12 -- see Fig \ref{sfig:gnetworkovertime}).
With this, we can describe $G4$ as having a middle-steep hierarchy that was maintained by many ongoing events among pairs.  This differs from $G1$ and $G3$, where the high fraction of events with the dominant individual suggests that the hierarchy was maintained mostly by these events.

During phase 3, the area available to each group was changed during Pds 11 and 12 by moving the compartment borders that separated the groups. In Pd 11, $G1$\&$G4$ had a larger area and $G2$\&$G3$ a smaller area. Pd 12 had these sizes switched. These manipulations did not have a consistent effect on space use or group social structure metrics (Fig \ref{sfig:arenasize}).

\subsection*{Individual social rankings, changes over time, and body mass}
\begin{figure}
	\centering
	\includegraphics[width=\linewidth]{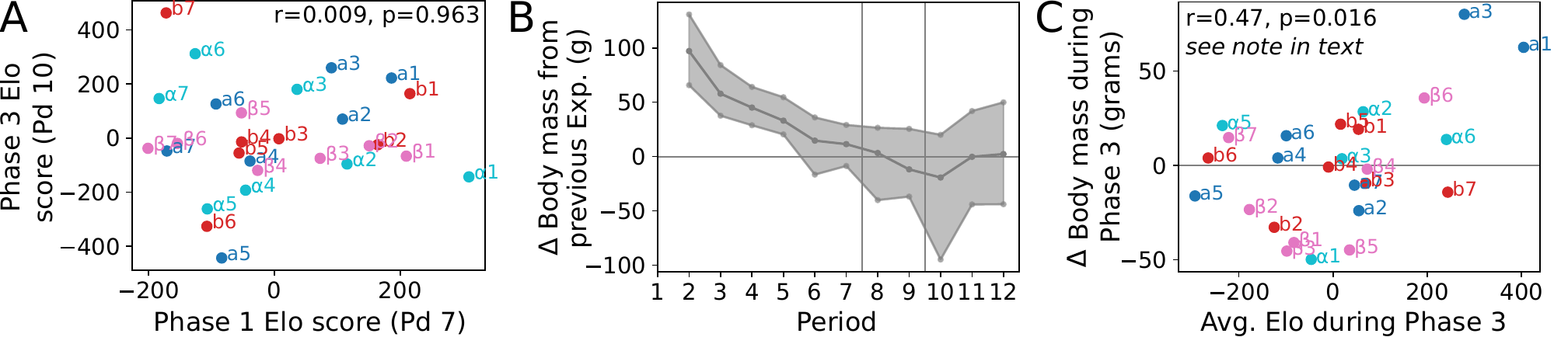}
	\caption{\textbf{Phase 1 to phase 3 individual rankings, body mass changes and social ranking}.
    (A) Comparison of individual rat ranking metrics at the end of phase 1 (Pd 7) to those at the start of phase 3 (Pd 10), after the new groups were formed.
    (B) Distribution of body mass changes over time for all individuals. The middle line is the mean, and rugged curves indicate the maximum/minimum across all individuals.  
    (C) Average Elo score during phase 3 compared to body mass change during phase 3 for each rat. Change in body mass during phase 3 is calculated as the body mass in Pd 12 minus body mass in Pd 10. 
	}
	\label{fig:bodymass}
\end{figure}

We found that previous social status in phase 1 did not predict an individual's placement in the new group social structures of phase 3 (Fig \ref{fig:bodymass}A).
This result holds if instead of using absolute Elo score values as shown in Fig \ref{fig:bodymass}A, rank scores of subordinate and dominant are used for the lowest two and highest two Elo scores, with other assigned as middle ranking (see Fig \ref{sfig:indivnetwork_p1_p3}).
Because the individual ranking metrics are correlated (Fig \ref{sfig:indivnetworkmetrics}), for clarity we focus on showing results the often-used Elo score \cite{albers_elo-rating_2001,neumann_assessing_2011,strauss_inferring_2019}; however, we also note that the other individual social metrics (including faction of events dominated, David's score, and local reaching centrality) showed similar trends (Fig \ref{sfig:indivnetwork_p1_p3}) with respect to predicting individual placement.


Because social ranking can be used to regulate access to resources such as food, we further examine the relationship between social rank and weight gain/loss.
We compare average individual social rankings to weight gain or loss during phase 3.
At the beginning of phase 1, all rats were young and gaining weight. However, by the end of phase 1, the average weight gain from the previous period was small, and not all rats were still gaining weight. When the groups were merged in phase 2, the average change in body mass ($\Delta$ body mass) continued to decrease and was negative for the last period of phase 2 and the first period of phase 3. Specifically, in the new groups of phase 3, the variance in the distribution of $\Delta$ body mass increased, with one rat (rat $\alpha 4$, which was subsequently permanently excluded from the experiment) losing nearly 100g relative to the previous period (Fig \ref{fig:bodymass}B).

Fig \ref{fig:bodymass}C shows that although the two most dominant rats during phase 3 gained the most weight (rats $a3$ and $a1$ from groups $G1$ and $G3$, respectively), average social dominance rank was not a robust general predictor of body mass changes across all rats.
Including these dominant rats, the relationship between the average Elo score during phase 3 and the change in body mass during phase 3 has a significant correlation with the values shown in Fig \ref{fig:bodymass}C.  However, this result is not robust: if these two rats are removed, the correlation drops greatly to $r=0.16$ and is no longer significant  ($p=0.44$). This result also holds if subordinate-middle-dominant ranks are used instead of absolute Elo score values ($r=0.48$ and $p=0.013$ including all individuals; $r=0.26$ and $p=0.23$ removing rats $a3$ and $a1$).
This demonstrates the complex relationship between dominance and body size in rodent social groups \cite{boreman_social_1972,macdonald_stability_1995}.
While there was likely a feedback regarding social rank and body mass for the two dominant rats in groups G1 and G3, respectively, it is difficult to link weight gain/loss to social rank in a general sense.

\subsection*{Individual metrics compared to behavioral assays}
We used individual and pairwise assays performed after the group experiments to test the behavior of each rat.
The individual black and white box, canopy, and elevated plus-maze results were used to define a composite boldness score.  
A pairwise social test with an unfamiliar rat, where two individuals are placed together and various behaviors characterizing interactions, such as sniffing the other, are scored (see Methods; Figure \ref{sfig:assays_pca}), was used to define a social interaction score.  This social test, which is also referred to in the literature as the ``reciprocal social interaction test'', has been widely employed for behavioral phenotyping related to anxiety and autism \cite{file_can_1978,file_review_2003,bolivar_assessing_2007,silverman_behavioural_2010}.

\begin{figure}
	\centering
		\includegraphics[width=\linewidth]{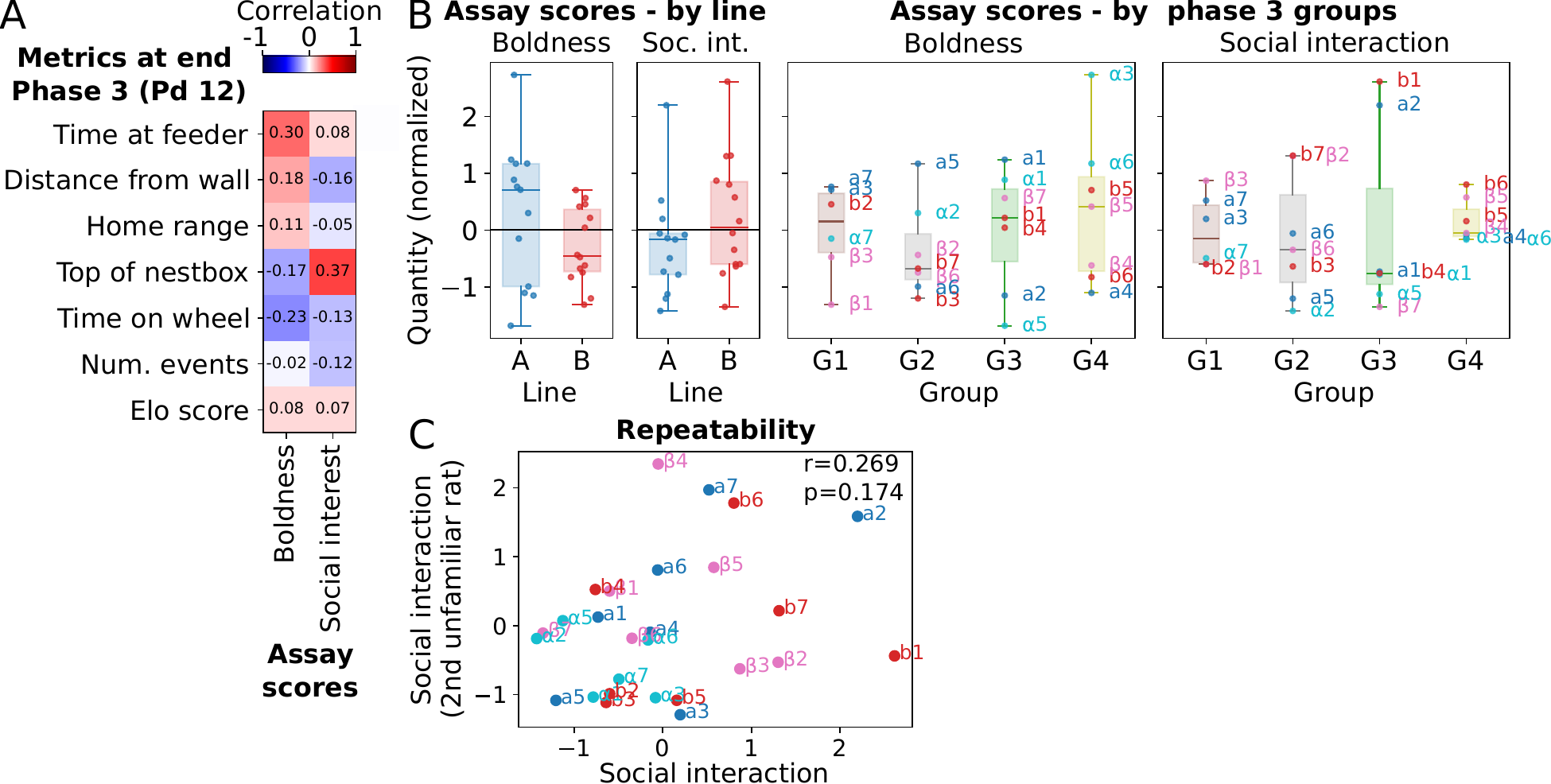}
	\caption{\textbf{Behavioral metrics at the end of phase 3 compared to assays}.
    (A) Pearson correlation values for space use and social behavioral metrics from the final period in phase 3 (Pd 12) with individual assay scores.  Labels and color scales denote correlation values.  Note that none of the correlations are significant (all p-values $>0.05$, calculated using t-distribution).
    (B) Individual score distributions according to breeding line (left), and by phase 3 group membership (right). Scores are normalized by the mean and standard deviation of values measured for all rats.	
        (C) Comparison of social interaction scores calculated from tests with a first unfamiliar rat (x-axes; values shown in A-B), with scores calculated from tests with a second unfamiliar rat (y-axes).  See also Figures \ref{sfig:indivtests_allscores} and \ref{sfig:assay_score_comparison}.
	}
	\label{fig:indivtests}
\end{figure}

In general, we find low and/or inconsistent relationships between behavior in groups and behavior in the assays (Figure \ref{fig:indivtests}A).
This is shown by the comparison of behavioral metrics from the last period in phase 3 with the individual boldness and social interaction scores.  
In particular, the social metrics measured in a group setting, including the number of events and the Elo score, do not exhibit consistent or significant correlations with the pairwise social interaction score.
Although we see positive correlations for the boldness score compared to the related metrics of distance from wall and home range, and a negative correlation with top of nestbox, these correlations are not significant ($p>0.05$), with a notable remark that the 2 most dominant individuals ($a3$ of $G1$, $a1$ of $G3$) have high boldness scores within their group.  
However, when considering the individual assays separately, we do find a significant correlation between top of nestbox and time spent in the open area during the elevated cross assay (Fig \ref{sfig:indivtests_allscores}A).
We also find that breeding line and group membership do not consistently predict differences in individual test scores (Figure \ref{fig:indivtests}B).  

The social interaction score shown in Figure \ref{fig:indivtests}A-B was obtained via pairwise behavioral assays performed with an unfamiliar rat.  
We repeated these tests with a second unfamiliar rat in order to test repeatability (tests were also performed with a familiar rat from the respective phase 3 groups - see Figure \ref{sfig:assay_score_comparison}).
The composite scores from the tests with the second unfamiliar individual show a low correlation with the scores obtained with the first unfamiliar individual (Figure \ref{fig:indivtests}C, $p>0.05$).

Other work has noted that individual behavior in assays may depend not only on an individual's social dominance status, but also on the nature of the social hierarchy of the group to which the individual previously belonged\cite{varholick_social_2019}. To test this, we fit a linear regression model predicting individual behavior assay results based on a combination of individual metrics and the Elo score steepness (ESS) of the group where the individual was located during Pd 12. While including this additional information increased explanatory power, we did not observe consistent significant patterns (Fig \ref{sfig:r2_metric_elo_steepness}).

\section*{Discussion}

We utilized automated tracking techniques to describe how rat groups develop and maintain a dynamic social structure over time, as well as how the social structure changes after regrouping.
Across successive periods, we observed a general consistency in the behavior of both individuals and groups. However, considering longer periods of time across multiple periods, we observed that the gradual accumulation of small changes can result in substantial behavioral changes over these longer time scales.
In addition, when the group composition was altered, we observed accelerated changes in behavior. 
Multiple metrics, including the Elo score, David's score, and reaching centrality, were employed to describe the overall hierarchical structure of each group as well as individual social rankings; these metrics revealed both similar and dissimilar aspects of the structure.
Different groups can vary a lot in their structure, and in particular, we found different and distinct social structures in the newly formed groups of phase 3.
We found that an individual's new position in the social hierarchy cannot be predicted based on their prior status when group composition changes.
While conventional individual assays (including the elevated cross, canopy, and other tests) produce consistent test results, we found that these measures have little correlation with individual behavior in a group setting.
Moreover, we found low repeatability in the scores measured with standard social test assays by performing the same test with different individuals; this also contributes to why behavior assays have little correlation with group behavior.

At the beginning of phase 1, the rats were still juveniles. Social interactions, particularly those related to aggression and dominance, are known to develop over time \cite{barnett_analysis_1958,calhoun_ecology_1963,bolles_ontogeny_1964}. Our observations are consistent with this, in particular, because we observed an increase in the number of approach-avoidance events and fights at the end of phase 1
(Figure \ref{sfig:indivmetrics}).
It is likely that in addition to group composition and interactions, the development stage also influenced the number of events and the differences in social structure from phase 1 to phase 3.
In a natural population, groups consist of individuals of different ages \cite{calhoun_ecology_1963}.
Targeted group mixing experiments -- for example, with both juveniles and adults in the same group -- could be used to ask how these effects interplay to generate overall emergent group social structures.

Behavioral assays are often used to quantify the behavior of rodents, and many new tools are being developed for both individual and social tracking and behavioral scoring \cite{spink_ethovision_2001,hong_automated_2015,de_chaumont_real-time_2019,nilsson_simple_2020,fong_pyrodenttracks_2022,szechtman_virtual_2022}. Tests that have been used to quantify social behavior in rodents include, for example, reciprocal social interaction, social approach partition, social preference, social transmission of food preference, food allocation, and reciprocal cooperation \cite{file_review_2003,schweinfurth_experimental_2017,schweinfurth_social_2020}. 
Most of these tests rely on creating artificial situations in order to measure the corresponding behavioral outcome. We compared social behavior measured in a group setting to scores of individuals measured with the reciprocal social interaction test. The pairwise social interaction test has known limitations, such as environmental dependencies, the possibility of aggression, and limited clarity as to which rat-initiated interactions \cite{kim_social_2019,acikgoz_overview_2022}. Although other behavior assays, such as the social preference test using a T-maze or modified home-cage tests, attempt to address limitations, these assays also have their own limitations \cite{kim_social_2019,acikgoz_overview_2022}.
It is an open question as to whether the social behavior measured in such tests can predict the social behavior under more natural conditions that include complex social interactions \cite{puscian_blueprints_2022}. 
In this respect, social interest/interaction and social dominance may represent different aspects of behavior, with the latter possibly only able to be measured in a group setting. Our comparison highlights the need for further work in this area.

We also note that while our group-living experiments provided space for complex environmental and social interactions, the conditions in the experiments were still much different from those that rats experience in the wild.  In this respect, our methods are similar to recent studies with mice, such as experiments with groups in the ``social box'' \cite{shemesh_high-order_2013,forkosh_identity_2019,karamihalev_social_2020}. These experimental setups can, therefore, be described as semi-natural.
A unique aspect of our study is the extended observation period, which allowed us to examine not only group social structure but also its change over time.

The study of social behavior is particularly important in animal models utilized for the understanding and treatment of social-related neuropsychiatric disorders. Rodents such as mice and rats have been indispensable as model biological organisms, with particular relevance to clinical research due to their short lifespan and tolerance to laboratory environments \cite{pearce_animal_2014,phifer-rixey_insights_2015,bale_critical_2019}. However, the laboratory environment typically restricts the development of complex social behaviors; for example, rats are often kept in small cages and then transferred to separate environments to examine social interactions using simplified behavioral assays like the 3-chambered social preference test \cite{silverman_behavioural_2010}. 
It is therefore not surprising that the translational relevance of these tests is limited \cite{homberg2013measuring,peters_ethological_2015}.  
Incorporating environmental and social complexity into experiments can increase the generalizability of conclusions drawn from laboratory studies \cite{kondrakiewicz_ecological_2019,shemesh_paradigm_2023}.
Moreover, long-term studies to examine group behavior may be an essential component to include in translational research applications, for example, in the testing of psychotherapeutic drugs to treat social anxiety \cite{homberg2013measuring,peters_ethological_2015}. In particular, an important topic for future work is to establish standardized and reproducible tests and measures that are properly representative of a full range of social behavior \cite{grieco_measuring_2021,shemesh_paradigm_2023}.
Furthermore, we note that much of our neuroscientific understanding of social behavior comes from dyadic interactions and reduced forms of social interactions \cite{kondrakiewicz_ecological_2019,gunaydin2014natural, padilla2022dynamic, lee2021neural}.
In light of the growing interest in the neuroscience of natural social behavior \cite{puscian_blueprints_2022}, 
going beyond basic social testing paradigms lends the opportunity to unravel a richer repertoire of neural mechanisms.

We note that while our social behavior analysis was used with video tracking data, it could be applied to other types of data, for example, data derived from markerless tracking methods, motion capture or QR-code tracking.
Future work can continue to expand on methods in this area, for example, including more detailed posture data, which can be used to describe social interactions in more detail \cite{mathis_deeplabcut_2018,graving_deepposekit_2019,pereira_fast_2019,pereira_sleap_2022}. Systems in this direction have already been developed for use with mice \cite{de_chaumont_real-time_2019}.
While we used {automated detection of} approach-avoidance interactions to define pairwise events, we note that a more detailed approach could use a combination of multiple behavioral interactions in order to define event ``winners'' and ``losers'' \cite{lehner_rats_2011}. {Moreover, detailed insight could be gained by using a hybrid method when the automated detection is followed by manual scoring of the behavior } \cite{pellis2022measuring,ham2023goldilocks}.
{Approach-avoidance as a measure of dominance has limitations, as the subordinate animal can freeze in place, with lack of movement signifying its subordinate status} \cite{fulenwider2022manifestations} {and this would be missed by the automated scoring scheme.}
Another area for future work is testing the functional consequences of group composition and social structure on individual or group performance, for example, with respect to search \cite{nagy_synergistic_2020}.

\section*{Author contributions}
T.V. and M.N. conceived the idea and designed the project.
M.N., G.V., D.Á., and E.K. conducted the experiments and collected the data.
E.K. performed the analysis of the individual assays.
G.V. designed and wrote the software framework for tracking the individuals.
M.N., G.V., and V.T. designed and performed the analysis of the behavior within the groups.
J.D.D., M.N., and A.E.H designed the analysis of the group-level data.
J.D.D. performed the analysis of the group-level data.
J.D.D., M.N., and A.E.H wrote the initial draft of the manuscript, and all authors revised the manuscript.

\section*{Acknowledgments}
We thank all the people who helped during the experimental work, especially Valéria Németh, Gergő Somorjai, Zsuzsa Ákos, Katalin Schlett, Péter Urtz, and András Péter (for Solomon coder). We acknowledge funding from Eötvös Loránd University and the Max Planck Institute of Animal Behavior. This research was partially supported by grants to T.V. establishing the MTA-ELTE Statistical and Biological Research Group, including NKFI Office Grant OTKA SNN 139598. Other sources of funding include the Hungarian Academy of Sciences (a grant to the MTA-ELTE ‘Lendület’ Collective Behaviour Research Group, grant number 95152, National Brain Programme 3.0 (NAP2022-I-3/2022), and the Companion Animal Research Group (PH1404/21), and the DFG Centre of Excellence 2117 ‘‘Centre for the Advanced Study of Collective Behaviour’’ (ID: 422037984). M.N. was partially supported by the Hungarian National Research, Development and Innovation Office (grant no. K 128780) and by Isaac Newton Institute for Mathematical Sciences, University of Cambridge. J.D. was partially supported by the Heidelberg Academy of Sciences Young Scientist (WIN-Kolleg) program.



\clearpage

\section*{Methods}

\subsection*{Resource availability}
\subsubsection*{Lead contact}
Further information and requests for resources and reagents should be directed to and will be fulfilled by the Lead Contact, Mate Nagy (nagymate@hal.elte.hu).
\subsubsection*{Materials availability} 
The study did not generate new unique reagents.

\subsubsection*{Data and code availability}
Trajectory and behavioral metrics data are available through Zenodo at \href{https://doi.org/10.5281/zenodo.7615468}{doi.org/10.5281/zenodo.7615468}. Recorded video sequences were analyzed offline with a custom-written software to obtain individual positions and orientations (source code available at \href{https://github.com/vasarhelyi/ratognize}{github.com/vasarhelyi/ratognize}), 
as well as trajectories and metrics (source code available at \href{https://github.com/vasarhelyi/trajognize}{github.com/vasarhelyi/trajognize}).
For details, see SI Methods of \cite{nagy_synergistic_2020}.
The scripts to run analyses included in this paper are available at \href{https://github.com/jacobdavidson/ratsocialgroups}{github.com/jacobdavidson/ratsocialgroups}.

\subsection*{Experimental model and subject details}
\subsubsection*{Subjects}
We used 28 Wistar male rats from 2 inbred breeding lines (14 Crl:WI BR and 14 HsdBrlHan:WIST, 7 litters/line, 2 individuals/litter; ordered from Toxi-Coop Zrt, Hungary) in this study. The rats arrived on 24 May 2011 at an age of 6 weeks. Rats were separated into 4 same-line groups (i.e.\ the phase 1 groups), each containing 7 rats from different litters, and were housed and tested together.

Each rat was marked with a unique 3-color barcode on its back using nontoxic ‘‘Special Effects’’ hair dye in 5 distinctive colors (Red: Nuclear Red, Orange: Napalm, Green: Sonic Green, Blue: Londa color 0/88, Purple: 4 units Atomic Pink and 1 unit Wild Flower). These codes were applied/renewed every 3 weeks.

\subsubsection*{Ethical guidelines}
The procedures comply with national and EU legislation and institutional guidelines. The experiments were performed in the animal facilities of Eötvös University, Hungary, and in accordance with Hungarian legislation and the corresponding definition by law (1998. évi XXVIII. Törvény 3. §/9. — the Animal Protection Act), which states that noninvasive studies on animals bred for research are allowed to be performed without the requirement of any special permission (PE/EA/1360-5/2018).

\subsection*{Method details}
\subsubsection*{Experimental conditions and monitoring}
Animals were housed in 4 compartments (sized 100 x 125 x 100 cm$^3$) with polypropylene covered wooden walls and sawdust changed weekly on a tiled floor (see Fig \ref{sfig:exampleframe}). Their room was kept at a controlled temperature of 21 ± 2°C, and with controlled light conditions featuring a daily cycle with 13h/11h dark/light. The dark (active) period was from 6am to 7pm with illumination at floor level $\sim$3-4 lux; the light period followed from 7pm to 6am of the following day with illumination of 300 lux. 
We video recorded the compartment 24/7 using a low-light sensitive camera fixed to the ceiling (Sony HDR-AX2000, 2.9 $\times$ 1.8 m$^2$ field of view, 1920 $\times$ 1080 resolution, 25 fps de-interlaced).
Rats had \textit{ad libitum} access to water and a shelter (nestbox), and access to food based on a fixed schedule for automated feeding. The feeding schedule followed a weekly cycle: 3 days (Sat, Sun, Mon) access to food 3 times for 1 hour (at 6am, noon, and 6pm); 3 days (Tue, Wed, Thu) access to food 2 times for 1 hour (at 6am and 6pm); and 1 day (Fri) access to food \textit{ad libitum} between 6am and 7pm). The housing compartments were cleaned once a week. 

We measured the weight of each individual three times a week (Mon, Thu, Fri; between 5pm and 6pm) and inspected their health. Some rats had injuries, and When a rat had a larger wound, we temporarily removed it until it was recovered. This happened on one occasion: $\alpha 7$ was taken out from week 33 to week 36 of the experiment. One rat ($\alpha 4$) was permanently removed from the experiment due to weight loss at week 31 at the age of 37 weeks.

\subsubsection*{Individual and social interaction tests}
These tests were performed at the end of the group measurement period, and included a total of 27 male rats at the age of 44 weeks. Subjects completed a test battery consisting of seven subtests examining fear-related and social behaviors in the following order (see descriptions below): black and white box, canopy, 
elevated plus-maze, 
social interaction test with outgroup conspecific, and social interaction test with in-group conspecific. The illumination during these tests was set according to the dark period mentioned above. 
Body weights were 480±70g (mean±SD) at the time of these tests.
The behavioral tests were conducted on three consecutive days between 10 a.m. and 6 p.m. (26 to 28 Feb 2012). The apparatuses were constructed from plastic sheets and cleaned between tests.
Depending on the test, either automated analysis was used to obtain trajectories or the behavior was coded by observers using the Solomon Coder software (beta 19.08.02). 
To ensure inter-observer reliability, pairs of observers overlapped in 20\% of the behavioral tests they scored. 
Using this overlap, the inter-observer reliability was calculated using the intraclass correlation coefficient (ICC) for all variables except the video frame variables, and we found all observations to be reliable (ICC $>$ 0.9).

\textbf{1. Black and white box}.
As described in Ramos et al.~\cite{ramos1997multiple}, the apparatus had a black and white compartment (each sized 27x27x27). The white compartment was strongly illuminated by a white bulb ($\sim$825 lux), while the black compartment was illuminated with a red bulb ($\sim$90 lux; Figure \ref{sfig:indiv_tests}A). The bulbs were positioned 37 cm above the apparatus floor. Each rat was initially placed in the center of the white compartment in a direction facing the opposing black compartment, and behavior was then recorded for 5 minutes. We tallied the number of video frames when the animal was in the 1) white compartment, 2) black compartment, or 3) at the border of the two areas and thus could not be clearly assigned (labeled as ``both'' in the data).

\textbf{2. Canopy}.
The apparatus consisted of a circular platform (104 cm in diameter) and a canopy (semitransparent red Plexiglas of 70 cm diameter) 10 cm above the platform (Figure \ref{sfig:indiv_tests}B). The mean illumination was 90 lux under the canopy and 400 lux outside of the canopy. At the beginning of the test, the animal was placed under the canopy. The test lasted for 5 minutes. We counted the number of video frames when the animal was 1) under the canopy, 2) in the exposed zone.

\textbf{3. Elevated plus-maze}.
Based on Ramos et al.~\cite{ramos1997multiple}, the apparatus had four elevated arms (66 cm from the floor), 45 cm long and 10 cm wide (Figure \ref{sfig:indiv_tests}C).
Two closed arms enclosed by a 50 cm high wall were located on opposing sides, and two open arms on the other two sides; the wall structure led to different illumination, with 25 lux for the closed arms and 65 lux for the open arms.
The central platform (10$\times$ 10 cm) connected the four arms to allow access to any. Each rat was first placed in the central platform facing an open arm, and subsequently, behavior was recorded for 5 minutes. 
To describe behavior, we counted the number of video frames in which the animal was in the 1) closed arms, 2) open arms, and 3) central platform (labeled as ``both'' in the data).



\textbf{4. Social interaction test with an unfamiliar (out-group) conspecific}.
In an open field arena, we placed an unfamiliar adult male next to a focal rat that had been part of the long-term experiment.  Two different unfamiliar rats were used for each phase 3 group (i.e.\ $G1$ rats were tested with unfamiliar rats 1 and 2, $G2$ rats with unfamiliar rats 3 and 4, etc.).
The test apparatus was made out of glass, with a green floor of 74 $\times$ 74 cm and transparent walls ($\sim$40 cm high; Figure \ref{sfig:indiv_tests}D).
The unfamiliar rats were significantly younger and smaller than the focals (12-weeks-old and 360±20g (mean±SD)). We recorded the behavior for 10 minutes. The test was repeated with other rats (unfamiliar and familiar) after a break that lasted for on average 75±42 min.\ but a minimum of 35 min. The coded behavior included the following: duration of bipedal orienting stance (\%), duration of self-grooming (\%), duration of exploration, duration of sniffing non-genital body parts (\%), duration of sniffing the genitals of the partner (\%), number of steps on the partner, number of fights. The coded parameters were stored separately from the trials with unfamiliar rats 1 and 2 for each focal rat.

\textbf{5. Social interaction test with a familiar (in-group) conspecific}.
The social interaction was also performed with a familiar conspecific, chosen as a random groupmate from their phase 3 group. Each individual was measured with four randomly selected members from their group. The coded variables were the same as above, but they were averaged over the trials for each focal rat.

\subsection*{Quantification and statistical analysis}
\subsubsection*{Data processing and behavioral metrics}
We calculated metrics of space use from the individual trajectory data for each rat.  Time at feeder is the fraction of total time spent at the feeder during nightlight (active period).  Distance from wall is the average distance from the walls during nightlight.  Home range is the area of an individual's space-use heatmap during all times (number of bins where it was detected more than 10 frames per day, calculated using bins with a linear size of $\sim$ 2 mm) (see Figure \ref{fig:illustrate}C),
normalized by total number of bins and frames counted.  Top of nestbox is the fraction of total time spent on top of the nestbox/shelter area during nightlight.

An approach-avoidance (AA) event was defined for a given pair of individuals ($i \ne j$) if, for a 0.4s long time window, the time-averaged dot product of $i$'s velocity ($v_i(t)$) and the normalized relative direction vector pointing from $i$ to $j$ 
-- a unit vector $\hat{d}_{ij}(t)={d}_{ij}(t)/|{d}_{ij}(t)|$, where
${d}_{ij}(t)=x_j(t)-x_i(t)$ is the relative position -- were within predetermined thresholds for both individuals.  The thresholds used were $AA_{ij}=\braket{(v_i(t) \cdot \hat{d}_{ji}(t)}_t>0.8$ for the approacher, and $AA_{ji}<-0.5$ for the avoider. In addition, we used the requirement that $i$ and $j$ were within 40 cm of each other ($|d_{ij}(t)| \leq d_{max} =40$ cm) and both $i$ and $j$ were moving at speeds of at least 0.25 m/sec ($|v(t)| \geq v_{min} =0.25$ m/sec). 

We use the approach-avoidance event network to quantify the social interaction structure in each group.  The values $A_{ij}$ are the number of times rat $i$ dominated approach-avoidance events with rat $j$.  Using this, the number of events rat $i$ dominated is $w_i = \sum_{j}A_{ij}$, and the number of events lost is $l_i = \sum_{j}A_{ji}$.  The fraction of events dominated for rat $i$ is then
\begin{equation}
    f_i = \frac{w_i}{w_i+l_i}.
\end{equation}

Reaching centrality is calculated using the normalized network of excess wins, $W_{ij}$.  This network has positive entries for the rat in a pair that dominated in more events and zero for the other rat, and is determined as
\begin{equation}
    W_{ij} = \frac{1}{Z}max\left(A_{ij}-A_{ji}, 0\right),
    \label{eq:Wij}
\end{equation}
where $Z$ is a normalization factor, which we define so that the maximum entry of $W_{ij}$ is equal to 1.
This network is then provided as input to the \verb|networkx| function \verb|local_reaching_centrality| to calculate the local reaching centrality (LRC) for each individual, and to the function \verb|global_reaching_centrality| to calculate global reaching centrality (GRC) for the group.  Note that we set the flag \verb|normalized=False| for calling these functions, because we use the definition in Eq \ref{eq:Wij} where $W_{ij}$ is already normalized.  With this, the LRC and GRC scores are in the range of 0 to 1.

We used the \verb|EloRating| package \cite{neumann_elochoice_2019} to calculate the individual David's score, David's score steepness (DSS), directional consistency index (DCI), and triangle transitivity index (TTRI).
For the individual Elo scores, we used the \verb|EloChoice| package \cite{neumann_elochoice_2019}, which has an improved and more efficient implementation of the randomization of interaction sequences used to calculate the Elo score.
We used the \verb|EloSteepness| package \cite{neumann_elosteepness_2023} to calculate the Elo score steepness (ESS).  These \textit{R} packages were integrated into our Python-based analysis code using \verb|rpy2|.


\subsubsection*{Individual and social interaction assays}
We used principal component analysis (PCA) to define the boldness and social interaction scores for each rat from the individual and pairwise assays.

\textbf{Boldness score}. 
The 8 videoframe variables calculated from the individual tests (black-and-white box, elevated plus-maze, and canopy test) were used to define the boldness score, which reflects the time spent in exposed portions of an unfamiliar environment.
We converted each frame count to a fraction of the test time and normalized the input variables before applying PCA. The first component explains 44.7\% of the variance, and positive projections onto this component represent more time in open areas (Figure \ref{sfig:assays_pca}A).
We used the projection of each rat onto the first component as the ``boldness'' score.
For comparison, we also calculated fractions of open-area time for each test:
fraction in the white area during black and white black box test (BWB-whitefrac = BWB-White/(BWB-White + BWB-Black)), fraction of time in open during the elevated cross test (ElevX-openfrac = ElevX-Open/(ElevX-Open + ElevX-Closed)), and fraction of time out during the canopy test (Canopy-outfrac = Canopy-OUT/(Canopy-Out + Canopy-Under)).

\textbf{Social interaction score}.
We applied PCA to measures from the pairwise social interaction tests and used results to define a composite score related to social interaction, and additionally used the 2nd PCA component to compare a ``self grooming" score.  The variables included are
duration of sniffing genitals (\%), 
duration of sniffing nongenital body parts (\%),
duration of bipedal orienting stance (standing up) (\%),
number of steps on the partner,
number of mating attempts,
number of fights,
duration of exploration (\%),
and duration of self grooming (\%).
The components shown in Figure \ref{sfig:assays_pca}B were determined using data from the first test with an unfamiliar rat; the scores for the other tests with a familiar rat and a second unfamiliar rat were calculated by projecting the associated variables onto these components.
The first PCA component represents interaction with the other rat, with positive projections indicating more interactions.  We use this component as the ``social interaction'' score.  The second PCA component is weighted most strongly by self grooming (positive) and exploration (negative) -- we addtionally compare this component as the ``self grooming'' score.

Figure \ref{sfig:assay_score_comparison} shows a comparison of the scores, including the boldness score and measures from the individual tests, and the social interaction and self grooming scores from the first test with an unfamiliar rat as well as tests with a familiar rat and a second unfamiliar rat.

\bibliography{references}

\begin{thebibliography}{10}
\urlstyle{rm}
\expandafter\ifx\csname url\endcsname\relax
  \def\url#1{\texttt{#1}}\fi
\expandafter\ifx\csname urlprefix\endcsname\relax\def\urlprefix{URL }\fi
\expandafter\ifx\csname doiprefix\endcsname\relax\def\doiprefix{DOI: }\fi
\providecommand{\bibinfo}[2]{#2}
\providecommand{\eprint}[2][]{\url{#2}}

\bibitem{vishwakarma_mechanical_2018}
\bibinfo{author}{Vishwakarma, M.} \emph{et~al.}
\newblock \bibinfo{journal}{\bibinfo{title}{Mechanical interactions among
  followers determine the emergence of leaders in migrating epithelial cell
  collectives}}.
\newblock {\emph{\JournalTitle{Nature Communications}}}
  \textbf{\bibinfo{volume}{9}}, \bibinfo{pages}{3469},
  \doiprefix\url{10.1038/s41467-018-05927-6} (\bibinfo{year}{2018}).

\bibitem{davidson_effect_2016}
\bibinfo{author}{Davidson, J.~D.}, \bibinfo{author}{Arauco-Aliaga, R.~P.},
  \bibinfo{author}{Crow, S.}, \bibinfo{author}{Gordon, D.~M.} \&
  \bibinfo{author}{Goldman, M.~S.}
\newblock \bibinfo{journal}{\bibinfo{title}{Effect of interactions between
  harvester ants on forager decisions}}.
\newblock {\emph{\JournalTitle{Frontiers in Ecology and Evolution}}}
  \textbf{\bibinfo{volume}{4}}, \bibinfo{pages}{115},
  \doiprefix\url{10.3389/fevo.2016.00115} (\bibinfo{year}{2016}).

\bibitem{nagy_synergistic_2020}
\bibinfo{author}{Nagy, M.} \emph{et~al.}
\newblock \bibinfo{journal}{\bibinfo{title}{Synergistic {Benefits} of {Group}
  {Search} in {Rats}}}.
\newblock {\emph{\JournalTitle{Current Biology}}}
  \textbf{\bibinfo{volume}{30}}, \bibinfo{pages}{4733--4738.e4},
  \doiprefix\url{10.1016/j.cub.2020.08.079} (\bibinfo{year}{2020}).

\bibitem{bond_61-million-person_2012}
\bibinfo{author}{Bond, R.~M.} \emph{et~al.}
\newblock \bibinfo{journal}{\bibinfo{title}{A 61-million-person experiment in
  social influence and political mobilization}}.
\newblock {\emph{\JournalTitle{Nature}}} \textbf{\bibinfo{volume}{489}},
  \bibinfo{pages}{295--298}, \doiprefix\url{10.1038/nature11421}
  (\bibinfo{year}{2012}).
\newblock \bibinfo{note}{Number: 7415 Publisher: Nature Publishing Group}.

\bibitem{kaburu_egalitarian_2015}
\bibinfo{author}{Kaburu, S. S.~K.} \& \bibinfo{author}{Newton-Fisher, N.~E.}
\newblock \bibinfo{journal}{\bibinfo{title}{Egalitarian despots: hierarchy
  steepness, reciprocity and the grooming-trade model in wild chimpanzees,
  {Pan} troglodytes}}.
\newblock {\emph{\JournalTitle{Animal Behaviour}}}
  \textbf{\bibinfo{volume}{99}}, \bibinfo{pages}{61--71},
  \doiprefix\url{10.1016/j.anbehav.2014.10.018} (\bibinfo{year}{2015}).

\bibitem{nagy_context-dependent_2013}
\bibinfo{author}{Nagy, M.} \emph{et~al.}
\newblock \bibinfo{journal}{\bibinfo{title}{Context-dependent hierarchies in
  pigeons}}.
\newblock {\emph{\JournalTitle{Proceedings of the National Academy of
  Sciences}}} \textbf{\bibinfo{volume}{110}}, \bibinfo{pages}{13049--13054},
  \doiprefix\url{10.1073/pnas.1305552110} (\bibinfo{year}{2013}).

\bibitem{maruska_social_2014}
\bibinfo{author}{Maruska, K.~P.}
\newblock \bibinfo{journal}{\bibinfo{title}{Social regulation of reproduction
  in male cichlid fishes}}.
\newblock {\emph{\JournalTitle{General and Comparative Endocrinology}}}
  \textbf{\bibinfo{volume}{207}}, \bibinfo{pages}{2--12},
  \doiprefix\url{10.1016/j.ygcen.2014.04.038} (\bibinfo{year}{2014}).

\bibitem{king_dominance_2008}
\bibinfo{author}{King, A.~J.}, \bibinfo{author}{Douglas, C.~M.},
  \bibinfo{author}{Huchard, E.}, \bibinfo{author}{Isaac, N.~J.} \&
  \bibinfo{author}{Cowlishaw, G.}
\newblock \bibinfo{journal}{\bibinfo{title}{Dominance and affiliation mediate
  despotism in a social primate}}.
\newblock {\emph{\JournalTitle{Current Biology}}}
  \textbf{\bibinfo{volume}{18}}, \bibinfo{pages}{1833--1838}
  (\bibinfo{year}{2008}).
\newblock \bibinfo{note}{Publisher: Elsevier}.

\bibitem{stewart_information_2019}
\bibinfo{author}{Stewart, A.~J.} \emph{et~al.}
\newblock \bibinfo{journal}{\bibinfo{title}{Information gerrymandering and
  undemocratic decisions}}.
\newblock {\emph{\JournalTitle{Nature}}} \textbf{\bibinfo{volume}{573}},
  \bibinfo{pages}{117--121}, \doiprefix\url{10.1038/s41586-019-1507-6}
  (\bibinfo{year}{2019}).
\newblock \bibinfo{note}{Number: 7772 Publisher: Nature Publishing Group}.

\bibitem{majolo_fitness-related_2012}
\bibinfo{author}{Majolo, B.}, \bibinfo{author}{Lehmann, J.},
  \bibinfo{author}{de~Bortoli~Vizioli, A.} \& \bibinfo{author}{Schino, G.}
\newblock \bibinfo{journal}{\bibinfo{title}{Fitness-related benefits of
  dominance in primates}}.
\newblock {\emph{\JournalTitle{American Journal of Physical Anthropology}}}
  \textbf{\bibinfo{volume}{147}}, \bibinfo{pages}{652--660},
  \doiprefix\url{10.1002/ajpa.22031} (\bibinfo{year}{2012}).
\newblock \bibinfo{note}{\_eprint:
  https://onlinelibrary.wiley.com/doi/pdf/10.1002/ajpa.22031}.

\bibitem{barnett_analysis_1958}
\bibinfo{author}{Barnett, S.~A.}
\newblock \bibinfo{journal}{\bibinfo{title}{An {Analysis} of {Social}
  {Behaviour} in {Wild} {Rats}}}.
\newblock {\emph{\JournalTitle{Proceedings of the Zoological Society of
  London}}} \textbf{\bibinfo{volume}{130}}, \bibinfo{pages}{107--152},
  \doiprefix\url{10.1111/j.1096-3642.1958.tb00565.x} (\bibinfo{year}{1958}).
\newblock \bibinfo{note}{\_eprint:
  https://onlinelibrary.wiley.com/doi/pdf/10.1111/j.1096-3642.1958.tb00565.x}.

\bibitem{grant_analysis_1963}
\bibinfo{author}{Grant, E.~C.}
\newblock \bibinfo{journal}{\bibinfo{title}{An {Analysis} of the {Social}
  {Behaviour} of the {Male} {Laboratory} {Rat}}}.
\newblock {\emph{\JournalTitle{Behaviour}}} \textbf{\bibinfo{volume}{21}},
  \bibinfo{pages}{260--281}, \doiprefix\url{10.1163/156853963X00194}
  (\bibinfo{year}{1963}).
\newblock \bibinfo{note}{Publisher: Brill}.

\bibitem{schweinfurth_social_2020}
\bibinfo{author}{Schweinfurth, M.~K.}
\newblock \bibinfo{journal}{\bibinfo{title}{The social life of {Norway} rats
  ({Rattus} norvegicus)}}.
\newblock {\emph{\JournalTitle{eLife}}} \textbf{\bibinfo{volume}{9}},
  \bibinfo{pages}{e54020}, \doiprefix\url{10.7554/eLife.54020}
  (\bibinfo{year}{2020}).
\newblock \bibinfo{note}{Publisher: eLife Sciences Publications, Ltd}.

\bibitem{dennis_systems_2021}
\bibinfo{author}{Dennis, E.~J.} \emph{et~al.}
\newblock \bibinfo{journal}{\bibinfo{title}{Systems {Neuroscience} of {Natural}
  {Behaviors} in {Rodents}}}.
\newblock {\emph{\JournalTitle{Journal of Neuroscience}}}
  \textbf{\bibinfo{volume}{41}}, \bibinfo{pages}{911--919}
  (\bibinfo{year}{2021}).

\bibitem{forkosh_animal_2021}
\bibinfo{author}{Forkosh, O.}
\newblock \bibinfo{journal}{\bibinfo{title}{Animal behavior and animal
  personality from a non-human perspective: {Getting} help from the machine}}.
\newblock {\emph{\JournalTitle{Patterns}}} \textbf{\bibinfo{volume}{2}},
  \doiprefix\url{10.1016/j.patter.2020.100194} (\bibinfo{year}{2021}).
\newblock \bibinfo{note}{Publisher: Elsevier}.

\bibitem{hinz_ontogeny_2017}
\bibinfo{author}{Hinz, R.~C.} \& \bibinfo{author}{Polavieja, G. G.~d.}
\newblock \bibinfo{journal}{\bibinfo{title}{Ontogeny of collective behavior
  reveals a simple attraction rule}}.
\newblock {\emph{\JournalTitle{Proceedings of the National Academy of
  Sciences}}} \bibinfo{pages}{201616926},
  \doiprefix\url{10.1073/pnas.1616926114} (\bibinfo{year}{2017}).

\bibitem{smith_behavioral_2022}
\bibinfo{author}{Smith, M.~L.} \emph{et~al.}
\newblock \bibinfo{journal}{\bibinfo{title}{Behavioral variation across the
  days and lives of honey bees}}.
\newblock {\emph{\JournalTitle{iScience}}} \textbf{\bibinfo{volume}{25}},
  \doiprefix\url{10.1016/j.isci.2022.104842} (\bibinfo{year}{2022}).

\bibitem{evans_long-term_2021}
\bibinfo{author}{Evans, J.~C.}, \bibinfo{author}{Lindholm, A.~K.} \&
  \bibinfo{author}{König, B.}
\newblock \bibinfo{journal}{\bibinfo{title}{Long-term overlap of social and
  genetic structure in free-ranging house mice reveals dynamic seasonal and
  group size effects}}.
\newblock {\emph{\JournalTitle{Current Zoology}}}
  \textbf{\bibinfo{volume}{67}}, \bibinfo{pages}{59--69},
  \doiprefix\url{10.1093/cz/zoaa030} (\bibinfo{year}{2021}).

\bibitem{jabarin_beyond_2022}
\bibinfo{author}{Jabarin, R.}, \bibinfo{author}{Netser, S.} \&
  \bibinfo{author}{Wagner, S.}
\newblock \bibinfo{journal}{\bibinfo{title}{Beyond the three-chamber test:
  toward a multimodal and objective assessment of social behavior in rodents}}.
\newblock {\emph{\JournalTitle{Molecular Autism}}}
  \textbf{\bibinfo{volume}{13}}, \bibinfo{pages}{41},
  \doiprefix\url{10.1186/s13229-022-00521-6} (\bibinfo{year}{2022}).

\bibitem{shemesh_high-order_2013}
\bibinfo{author}{Shemesh, Y.} \emph{et~al.}
\newblock \bibinfo{journal}{\bibinfo{title}{High-order social interactions in
  groups of mice}}.
\newblock {\emph{\JournalTitle{eLife}}} \textbf{\bibinfo{volume}{2}},
  \bibinfo{pages}{e00759}, \doiprefix\url{10.7554/eLife.00759}
  (\bibinfo{year}{2013}).

\bibitem{forkosh_identity_2019}
\bibinfo{author}{Forkosh, O.} \emph{et~al.}
\newblock \bibinfo{journal}{\bibinfo{title}{Identity domains capture individual
  differences from across the behavioral repertoire}}.
\newblock {\emph{\JournalTitle{Nature Neuroscience}}}
  \textbf{\bibinfo{volume}{22}}, \bibinfo{pages}{2023--2028},
  \doiprefix\url{10.1038/s41593-019-0516-y} (\bibinfo{year}{2019}).
\newblock \bibinfo{note}{Number: 12 Publisher: Nature Publishing Group}.

\bibitem{karamihalev_social_2020}
\bibinfo{author}{Karamihalev, S.} \emph{et~al.}
\newblock \bibinfo{journal}{\bibinfo{title}{Social dominance mediates
  behavioral adaptation to chronic stress in a sex-specific manner}}.
\newblock {\emph{\JournalTitle{eLife}}} \textbf{\bibinfo{volume}{9}},
  \bibinfo{pages}{e58723}, \doiprefix\url{10.7554/eLife.58723}
  (\bibinfo{year}{2020}).
\newblock \bibinfo{note}{Publisher: eLife Sciences Publications, Ltd}.

\bibitem{lopez_ketamine_2022}
\bibinfo{author}{Lopez, J.~P.} \emph{et~al.}
\newblock \bibinfo{journal}{\bibinfo{title}{Ketamine exerts its sustained
  antidepressant effects via cell-type-specific regulation of {Kcnq2}}}.
\newblock {\emph{\JournalTitle{Neuron}}} \textbf{\bibinfo{volume}{110}},
  \bibinfo{pages}{2283--2298.e9}, \doiprefix\url{10.1016/j.neuron.2022.05.001}
  (\bibinfo{year}{2022}).

\bibitem{williamson_temporal_2016}
\bibinfo{author}{Williamson, C.~M.}, \bibinfo{author}{Lee, W.} \&
  \bibinfo{author}{Curley, J.~P.}
\newblock \bibinfo{journal}{\bibinfo{title}{Temporal dynamics of social
  hierarchy formation and maintenance in male mice}}.
\newblock {\emph{\JournalTitle{Animal Behaviour}}}
  \textbf{\bibinfo{volume}{115}}, \bibinfo{pages}{259--272},
  \doiprefix\url{10.1016/j.anbehav.2016.03.004} (\bibinfo{year}{2016}).

\bibitem{lee_effect_2021}
\bibinfo{author}{Lee, W.} \emph{et~al.}
\newblock \bibinfo{journal}{\bibinfo{title}{Effect of relative social rank
  within a social hierarchy on neural activation in response to familiar or
  unfamiliar social signals}}.
\newblock {\emph{\JournalTitle{Scientific Reports}}}
  \textbf{\bibinfo{volume}{11}}, \bibinfo{pages}{2864},
  \doiprefix\url{10.1038/s41598-021-82255-8} (\bibinfo{year}{2021}).
\newblock \bibinfo{note}{Number: 1 Publisher: Nature Publishing Group}.

\bibitem{lee_distinct_2022}
\bibinfo{author}{Lee, W.} \emph{et~al.}
\newblock \bibinfo{journal}{\bibinfo{title}{Distinct immune and transcriptomic
  profiles in dominant versus subordinate males in mouse social hierarchies}}.
\newblock {\emph{\JournalTitle{Brain, Behavior, and Immunity}}}
  \textbf{\bibinfo{volume}{103}}, \bibinfo{pages}{130--144},
  \doiprefix\url{10.1016/j.bbi.2022.04.015} (\bibinfo{year}{2022}).

\bibitem{johnson_supplanting_1989}
\bibinfo{author}{Johnson, J.~A.}
\newblock \bibinfo{journal}{\bibinfo{title}{Supplanting by olive baboons:
  dominance rank difference and resource value}}.
\newblock {\emph{\JournalTitle{Behavioral Ecology and Sociobiology}}}
  \textbf{\bibinfo{volume}{24}}, \bibinfo{pages}{277--283},
  \doiprefix\url{10.1007/BF00290903} (\bibinfo{year}{1989}).

\bibitem{evans_inferring_2018}
\bibinfo{author}{Evans, J.~C.}, \bibinfo{author}{Devost, I.},
  \bibinfo{author}{Jones, T.~B.} \& \bibinfo{author}{Morand-Ferron, J.}
\newblock \bibinfo{journal}{\bibinfo{title}{Inferring dominance interactions
  from automatically recorded temporal data}}.
\newblock {\emph{\JournalTitle{Ethology}}} \textbf{\bibinfo{volume}{124}},
  \bibinfo{pages}{188--195}, \doiprefix\url{10.1111/eth.12720}
  (\bibinfo{year}{2018}).

\bibitem{gullstrand_computerized_2021}
\bibinfo{author}{Gullstrand, J.}, \bibinfo{author}{Claidière, N.} \&
  \bibinfo{author}{Fagot, J.}
\newblock \bibinfo{journal}{\bibinfo{title}{Computerized assessment of
  dominance hierarchy in baboons ({Papio} papio)}}.
\newblock {\emph{\JournalTitle{Behavior Research Methods}}}
  \textbf{\bibinfo{volume}{53}}, \bibinfo{pages}{1923--1934},
  \doiprefix\url{10.3758/s13428-021-01539-z} (\bibinfo{year}{2021}).

\bibitem{spruijt_approach_1992}
\bibinfo{author}{Spruijt, B.~M.}, \bibinfo{author}{Hol, T.} \&
  \bibinfo{author}{Rousseau, J.}
\newblock \bibinfo{journal}{\bibinfo{title}{Approach, avoidance, and contact
  behavior of individually recognized animals automatically quantified with an
  imaging technique}}.
\newblock {\emph{\JournalTitle{Physiology \& Behavior}}}
  \textbf{\bibinfo{volume}{51}}, \bibinfo{pages}{747--752},
  \doiprefix\url{10.1016/0031-9384(92)90111-E} (\bibinfo{year}{1992}).

\bibitem{elo_rating_1978}
\bibinfo{author}{Elo, A.~E.}
\newblock \emph{\bibinfo{title}{The rating of chessplayers, past and present}}
  (\bibinfo{publisher}{Arco Pub}, \bibinfo{address}{New York},
  \bibinfo{year}{1978}).

\bibitem{albers_elo-rating_2001}
\bibinfo{author}{Albers, P.~C.} \& \bibinfo{author}{De~Vries, H.}
\newblock \bibinfo{journal}{\bibinfo{title}{Elo-rating as a tool in the
  sequential estimation of dominance strengths}}.
\newblock {\emph{\JournalTitle{Animal Behaviour}}}
  \textbf{\bibinfo{volume}{61}}, \bibinfo{pages}{489--495},
  \doiprefix\url{10.1006/anbe.2000.1571} (\bibinfo{year}{2001}).

\bibitem{neumann_assessing_2011}
\bibinfo{author}{Neumann, C.} \emph{et~al.}
\newblock \bibinfo{journal}{\bibinfo{title}{Assessing dominance hierarchies:
  validation and advantages of progressive evaluation with {Elo}-rating}}.
\newblock {\emph{\JournalTitle{Animal Behaviour}}}
  \textbf{\bibinfo{volume}{82}}, \bibinfo{pages}{911--921},
  \doiprefix\url{10.1016/j.anbehav.2011.07.016} (\bibinfo{year}{2011}).

\bibitem{strauss_inferring_2019}
\bibinfo{author}{Strauss, E.~D.} \& \bibinfo{author}{Holekamp, K.~E.}
\newblock \bibinfo{journal}{\bibinfo{title}{Inferring longitudinal hierarchies:
  {Framework} and methods for studying the dynamics of dominance}}.
\newblock {\emph{\JournalTitle{Journal of Animal Ecology}}}
  \textbf{\bibinfo{volume}{88}}, \bibinfo{pages}{521--536},
  \doiprefix\url{10.1111/1365-2656.12951} (\bibinfo{year}{2019}).
\newblock \bibinfo{note}{\_eprint:
  https://onlinelibrary.wiley.com/doi/pdf/10.1111/1365-2656.12951}.

\bibitem{mones_hierarchy_2012}
\bibinfo{author}{Mones, E.}, \bibinfo{author}{Vicsek, L.} \&
  \bibinfo{author}{Vicsek, T.}
\newblock \bibinfo{journal}{\bibinfo{title}{Hierarchy {Measure} for {Complex}
  {Networks}}}.
\newblock {\emph{\JournalTitle{PLOS ONE}}} \textbf{\bibinfo{volume}{7}},
  \bibinfo{pages}{e33799}, \doiprefix\url{10.1371/journal.pone.0033799}
  (\bibinfo{year}{2012}).
\newblock \bibinfo{note}{Publisher: Public Library of Science}.

\bibitem{ozogany_modeling_2015}
\bibinfo{author}{Ozogány, K.} \& \bibinfo{author}{Vicsek, T.}
\newblock \bibinfo{journal}{\bibinfo{title}{Modeling the {Emergence} of
  {Modular} {Leadership} {Hierarchy} {During} the {Collective} {Motion} of
  {Herds} {Made} of {Harems}}}.
\newblock {\emph{\JournalTitle{Journal of Statistical Physics}}}
  \textbf{\bibinfo{volume}{158}}, \bibinfo{pages}{628--646},
  \doiprefix\url{10.1007/s10955-014-1131-7} (\bibinfo{year}{2015}).

\bibitem{shimoji_global_2014}
\bibinfo{author}{Shimoji, H.}, \bibinfo{author}{Abe, M.~S.},
  \bibinfo{author}{Tsuji, K.} \& \bibinfo{author}{Masuda, N.}
\newblock \bibinfo{journal}{\bibinfo{title}{Global network structure of
  dominance hierarchy of ant workers}}.
\newblock {\emph{\JournalTitle{Journal of The Royal Society Interface}}}
  \textbf{\bibinfo{volume}{11}}, \bibinfo{pages}{20140599},
  \doiprefix\url{10.1098/rsif.2014.0599} (\bibinfo{year}{2014}).
\newblock \bibinfo{note}{Publisher: Royal Society}.

\bibitem{kora_global_2023}
\bibinfo{author}{Kora, Y.}, \bibinfo{author}{Salhi, S.},
  \bibinfo{author}{Davidsen, J.} \& \bibinfo{author}{Simon, C.}
\newblock \bibinfo{journal}{\bibinfo{title}{Global excitability and network
  structure in the human brain}}.
\newblock {\emph{\JournalTitle{Physical Review E}}}
  \textbf{\bibinfo{volume}{107}}, \bibinfo{pages}{054308},
  \doiprefix\url{10.1103/PhysRevE.107.054308} (\bibinfo{year}{2023}).

\bibitem{hu_hierarchy_2017}
\bibinfo{author}{Hu, F.}, \bibinfo{author}{Zhao, S.}, \bibinfo{author}{Bing,
  T.} \& \bibinfo{author}{Chang, Y.}
\newblock \bibinfo{journal}{\bibinfo{title}{Hierarchy in industrial structure:
  {The} cases of {China} and the {USA}}}.
\newblock {\emph{\JournalTitle{Physica A: Statistical Mechanics and its
  Applications}}} \textbf{\bibinfo{volume}{469}}, \bibinfo{pages}{871--882},
  \doiprefix\url{10.1016/j.physa.2016.11.083} (\bibinfo{year}{2017}).

\bibitem{beardsley_hierarchy_2020}
\bibinfo{author}{Beardsley, K.}, \bibinfo{author}{Liu, H.},
  \bibinfo{author}{Mucha, P.~J.}, \bibinfo{author}{Siegel, D.~A.} \&
  \bibinfo{author}{Tellez, J.~F.}
\newblock \bibinfo{journal}{\bibinfo{title}{Hierarchy and the {Provision} of
  {Order} in {International} {Politics}}}.
\newblock {\emph{\JournalTitle{The Journal of Politics}}}
  \textbf{\bibinfo{volume}{82}}, \bibinfo{pages}{731--746},
  \doiprefix\url{10.1086/707096} (\bibinfo{year}{2020}).
\newblock \bibinfo{note}{Publisher: The University of Chicago Press}.

\bibitem{mones_universal_2014}
\bibinfo{author}{Mones, E.}, \bibinfo{author}{Pollner, P.} \&
  \bibinfo{author}{Vicsek, T.}
\newblock \bibinfo{journal}{\bibinfo{title}{Universal hierarchical behavior of
  citation networks}}.
\newblock {\emph{\JournalTitle{Journal of Statistical Mechanics: Theory and
  Experiment}}} \textbf{\bibinfo{volume}{2014}}, \bibinfo{pages}{P05023},
  \doiprefix\url{10.1088/1742-5468/2014/05/P05023} (\bibinfo{year}{2014}).
\newblock \bibinfo{note}{Publisher: IOP Publishing and SISSA}.

\bibitem{strauss_domarchive_2022}
\bibinfo{author}{Strauss, E.~D.} \emph{et~al.}
\newblock \bibinfo{journal}{\bibinfo{title}{{DomArchive}: a century of
  published dominance data}}.
\newblock {\emph{\JournalTitle{Philosophical Transactions of the Royal Society
  B: Biological Sciences}}} \textbf{\bibinfo{volume}{377}},
  \bibinfo{pages}{20200436}, \doiprefix\url{10.1098/rstb.2020.0436}
  (\bibinfo{year}{2022}).
\newblock \bibinfo{note}{Publisher: Royal Society}.

\bibitem{shizuka_social_2012}
\bibinfo{author}{Shizuka, D.} \& \bibinfo{author}{McDonald, D.~B.}
\newblock \bibinfo{journal}{\bibinfo{title}{A social network perspective on
  measurements of dominance hierarchies}}.
\newblock {\emph{\JournalTitle{Animal Behaviour}}}
  \textbf{\bibinfo{volume}{83}}, \bibinfo{pages}{925--934},
  \doiprefix\url{10.1016/j.anbehav.2012.01.011} (\bibinfo{year}{2012}).

\bibitem{de_vries_improved_1995}
\bibinfo{author}{de~Vries, H.}
\newblock \bibinfo{journal}{\bibinfo{title}{An improved test of linearity in
  dominance hierarchies containing unknown or tied relationships}}.
\newblock {\emph{\JournalTitle{Animal Behaviour}}}
  \textbf{\bibinfo{volume}{50}}, \bibinfo{pages}{1375--1389},
  \doiprefix\url{10.1016/0003-3472(95)80053-0} (\bibinfo{year}{1995}).

\bibitem{neumann_extending_2023}
\bibinfo{author}{Neumann, C.} \& \bibinfo{author}{Fischer, J.}
\newblock \bibinfo{journal}{\bibinfo{title}{Extending {Bayesian} {Elo}-rating
  to quantify the steepness of dominance hierarchies}}.
\newblock {\emph{\JournalTitle{Methods in Ecology and Evolution}}}
  \textbf{\bibinfo{volume}{14}}, \bibinfo{pages}{669--682},
  \doiprefix\url{10.1111/2041-210X.14021} (\bibinfo{year}{2023}).
\newblock \bibinfo{note}{\_eprint:
  https://onlinelibrary.wiley.com/doi/pdf/10.1111/2041-210X.14021}.

\bibitem{burt1943territoriality}
\bibinfo{author}{Burt, W.~H.}
\newblock \bibinfo{journal}{\bibinfo{title}{Territoriality and home range
  concepts as applied to mammals}}.
\newblock {\emph{\JournalTitle{Journal of mammalogy}}}
  \textbf{\bibinfo{volume}{24}}, \bibinfo{pages}{346--352}
  (\bibinfo{year}{1943}).

\bibitem{varholick_social_2019}
\bibinfo{author}{Varholick, J.~A.} \emph{et~al.}
\newblock \bibinfo{journal}{\bibinfo{title}{Social dominance hierarchy type and
  rank contribute to phenotypic variation within cages of laboratory mice}}.
\newblock {\emph{\JournalTitle{Scientific Reports}}}
  \textbf{\bibinfo{volume}{9}}, \bibinfo{pages}{13650},
  \doiprefix\url{10.1038/s41598-019-49612-0} (\bibinfo{year}{2019}).
\newblock \bibinfo{note}{Publisher: Nature Publishing Group}.

\bibitem{de_vries_measuring_2006}
\bibinfo{author}{de~Vries, H.}, \bibinfo{author}{Stevens, J. M.~G.} \&
  \bibinfo{author}{Vervaecke, H.}
\newblock \bibinfo{journal}{\bibinfo{title}{Measuring and testing the steepness
  of dominance hierarchies}}.
\newblock {\emph{\JournalTitle{Animal Behaviour}}}
  \textbf{\bibinfo{volume}{71}}, \bibinfo{pages}{585--592},
  \doiprefix\url{10.1016/j.anbehav.2005.05.015} (\bibinfo{year}{2006}).

\bibitem{van_hooff_dominance_1987}
\bibinfo{author}{van Hooff, J. A. R. A.~M.} \& \bibinfo{author}{Wensing, J.
  A.~B.}
\newblock \bibinfo{title}{Dominance and its behavioral measures in a captive
  wolf pack}.
\newblock In \emph{\bibinfo{booktitle}{Man and wolf: {Advances}, issues, and
  problems in captive wolf research}}, Perspectives in vertebrate science,
  {Vol}. 4., \bibinfo{pages}{219--252} (\bibinfo{publisher}{Dr W Junk
  Publishers}, \bibinfo{address}{Dordrecht, Netherlands},
  \bibinfo{year}{1987}).

\bibitem{shizuka_network_2015}
\bibinfo{author}{Shizuka, D.} \& \bibinfo{author}{McDonald, D.~B.}
\newblock \bibinfo{journal}{\bibinfo{title}{The network motif architecture of
  dominance hierarchies}}.
\newblock {\emph{\JournalTitle{Journal of The Royal Society Interface}}}
  \textbf{\bibinfo{volume}{12}}, \bibinfo{pages}{20150080},
  \doiprefix\url{10.1098/rsif.2015.0080} (\bibinfo{year}{2015}).
\newblock \bibinfo{note}{Publisher: Royal Society}.

\bibitem{boreman_social_1972}
\bibinfo{author}{Boreman, J.} \& \bibinfo{author}{Price, E.}
\newblock \bibinfo{journal}{\bibinfo{title}{Social dominance in wild and
  domestic {Norway} rats ({Rattus} norvegicus)}}.
\newblock {\emph{\JournalTitle{Animal Behaviour}}}
  \textbf{\bibinfo{volume}{20}}, \bibinfo{pages}{534--542},
  \doiprefix\url{10.1016/S0003-3472(72)80018-6} (\bibinfo{year}{1972}).

\bibitem{macdonald_stability_1995}
\bibinfo{author}{Macdonald, D.~W.}, \bibinfo{author}{Berdoy, M.} \&
  \bibinfo{author}{Smith, P.}
\newblock \bibinfo{journal}{\bibinfo{title}{Stability of {Social} {Status} in
  {Wild} {Rats}: {Age} and the {Role} of {Settled} {Dominance}}}.
\newblock {\emph{\JournalTitle{Behaviour}}} \textbf{\bibinfo{volume}{132}},
  \bibinfo{pages}{193--212}, \doiprefix\url{10.1163/156853995X00694}
  (\bibinfo{year}{1995}).
\newblock \bibinfo{note}{Publisher: Brill}.

\bibitem{file_can_1978}
\bibinfo{author}{File, S.~E.} \& \bibinfo{author}{Hyde, J.~R.}
\newblock \bibinfo{journal}{\bibinfo{title}{Can social interaction be used to
  measure anxiety?}}
\newblock {\emph{\JournalTitle{British Journal of Pharmacology}}}
  \textbf{\bibinfo{volume}{62}}, \bibinfo{pages}{19--24}
  (\bibinfo{year}{1978}).

\bibitem{file_review_2003}
\bibinfo{author}{File, S.~E.} \& \bibinfo{author}{Seth, P.}
\newblock \bibinfo{journal}{\bibinfo{title}{A review of 25 years of the social
  interaction test}}.
\newblock {\emph{\JournalTitle{European Journal of Pharmacology}}}
  \textbf{\bibinfo{volume}{463}}, \bibinfo{pages}{35--53},
  \doiprefix\url{10.1016/S0014-2999(03)01273-1} (\bibinfo{year}{2003}).

\bibitem{bolivar_assessing_2007}
\bibinfo{author}{Bolivar, V.~J.}, \bibinfo{author}{Walters, S.~R.} \&
  \bibinfo{author}{Phoenix, J.~L.}
\newblock \bibinfo{journal}{\bibinfo{title}{Assessing autism-like behavior in
  mice: {Variations} in social interactions among inbred strains}}.
\newblock {\emph{\JournalTitle{Behavioural Brain Research}}}
  \textbf{\bibinfo{volume}{176}}, \bibinfo{pages}{21--26},
  \doiprefix\url{10.1016/j.bbr.2006.09.007} (\bibinfo{year}{2007}).

\bibitem{silverman_behavioural_2010}
\bibinfo{author}{Silverman, J.~L.}, \bibinfo{author}{Yang, M.},
  \bibinfo{author}{Lord, C.} \& \bibinfo{author}{Crawley, J.~N.}
\newblock \bibinfo{journal}{\bibinfo{title}{Behavioural phenotyping assays for
  mouse models of autism}}.
\newblock {\emph{\JournalTitle{Nature Reviews Neuroscience}}}
  \textbf{\bibinfo{volume}{11}}, \bibinfo{pages}{490--502},
  \doiprefix\url{10.1038/nrn2851} (\bibinfo{year}{2010}).
\newblock \bibinfo{note}{Number: 7 Publisher: Nature Publishing Group}.

\bibitem{calhoun_ecology_1963}
\bibinfo{author}{Calhoun, J.~B.}
\newblock \emph{\bibinfo{title}{The {Ecology} and {Sociology} of the {Norway}
  {Rat}}} (\bibinfo{publisher}{U.S. Department of Health, Education, and
  Welfare, Public Health Service}, \bibinfo{year}{1963}).
\newblock \bibinfo{note}{Google-Books-ID: crhqAAAAMAAJ}.

\bibitem{bolles_ontogeny_1964}
\bibinfo{author}{Bolles, R.~C.} \& \bibinfo{author}{Woods, P.~J.}
\newblock \bibinfo{journal}{\bibinfo{title}{The ontogeny of behaviour in the
  albino rat}}.
\newblock {\emph{\JournalTitle{Animal Behaviour}}}
  \textbf{\bibinfo{volume}{12}}, \bibinfo{pages}{427--441},
  \doiprefix\url{10.1016/0003-3472(64)90062-4} (\bibinfo{year}{1964}).

\bibitem{spink_ethovision_2001}
\bibinfo{author}{Spink, A.~J.}, \bibinfo{author}{Tegelenbosch, R. A.~J.},
  \bibinfo{author}{Buma, M. O.~S.} \& \bibinfo{author}{Noldus, L. P. J.~J.}
\newblock \bibinfo{journal}{\bibinfo{title}{The {EthoVision} video tracking
  system—{A} tool for behavioral phenotyping of transgenic mice}}.
\newblock {\emph{\JournalTitle{Physiology \& Behavior}}}
  \textbf{\bibinfo{volume}{73}}, \bibinfo{pages}{731--744},
  \doiprefix\url{10.1016/S0031-9384(01)00530-3} (\bibinfo{year}{2001}).

\bibitem{hong_automated_2015}
\bibinfo{author}{Hong, W.} \emph{et~al.}
\newblock \bibinfo{journal}{\bibinfo{title}{Automated measurement of mouse
  social behaviors using depth sensing, video tracking, and machine learning}}.
\newblock {\emph{\JournalTitle{Proceedings of the National Academy of
  Sciences}}} \textbf{\bibinfo{volume}{112}}, \bibinfo{pages}{E5351--E5360},
  \doiprefix\url{10.1073/pnas.1515982112} (\bibinfo{year}{2015}).
\newblock \bibinfo{note}{Publisher: Proceedings of the National Academy of
  Sciences}.

\bibitem{de_chaumont_real-time_2019}
\bibinfo{author}{de~Chaumont, F.} \emph{et~al.}
\newblock \bibinfo{journal}{\bibinfo{title}{Real-time analysis of the behaviour
  of groups of mice via a depth-sensing camera and machine learning}}.
\newblock {\emph{\JournalTitle{Nature Biomedical Engineering}}}
  \textbf{\bibinfo{volume}{3}}, \bibinfo{pages}{930--942},
  \doiprefix\url{10.1038/s41551-019-0396-1} (\bibinfo{year}{2019}).
\newblock \bibinfo{note}{Number: 11 Publisher: Nature Publishing Group}.

\bibitem{nilsson_simple_2020}
\bibinfo{author}{Nilsson, S.~R.} \emph{et~al.}
\newblock \bibinfo{title}{Simple {Behavioral} {Analysis} ({SimBA}) – an open
  source toolkit for computer classification of complex social behaviors in
  experimental animals}, \doiprefix\url{10.1101/2020.04.19.049452}
  (\bibinfo{year}{2020}).
\newblock \bibinfo{note}{Pages: 2020.04.19.049452 Section: New Results}.

\bibitem{fong_pyrodenttracks_2022}
\bibinfo{author}{Fong, T.}, \bibinfo{author}{Jury, B.}, \bibinfo{author}{Hu,
  H.} \& \bibinfo{author}{Murphy, T.~H.}
\newblock \bibinfo{title}{{PyRodentTracks}: flexible computer vision and {RFID}
  based system for multiple rodent tracking and behavioral assessment},
  \doiprefix\url{10.1101/2022.01.23.477395} (\bibinfo{year}{2022}).
\newblock \bibinfo{note}{Pages: 2022.01.23.477395 Section: New Results}.

\bibitem{szechtman_virtual_2022}
\bibinfo{author}{Szechtman, H.}, \bibinfo{author}{Dvorkin-Gheva, A.} \&
  \bibinfo{author}{Gomez-Marin, A.}
\newblock \bibinfo{journal}{\bibinfo{title}{A virtual library for behavioral
  performance in standard conditions—rodent spontaneous activity in an open
  field during repeated testing and after treatment with drugs or brain
  lesions}}.
\newblock {\emph{\JournalTitle{GigaScience}}} \textbf{\bibinfo{volume}{11}},
  \bibinfo{pages}{giac092}, \doiprefix\url{10.1093/gigascience/giac092}
  (\bibinfo{year}{2022}).

\bibitem{schweinfurth_experimental_2017}
\bibinfo{author}{Schweinfurth, M.~K.}, \bibinfo{author}{Stieger, B.} \&
  \bibinfo{author}{Taborsky, M.}
\newblock \bibinfo{journal}{\bibinfo{title}{Experimental evidence for
  reciprocity in allogrooming among wild-type {Norway} rats}}.
\newblock {\emph{\JournalTitle{Scientific Reports}}}
  \textbf{\bibinfo{volume}{7}}, \bibinfo{pages}{4010},
  \doiprefix\url{10.1038/s41598-017-03841-3} (\bibinfo{year}{2017}).
\newblock \bibinfo{note}{Number: 1 Publisher: Nature Publishing Group}.

\bibitem{kim_social_2019}
\bibinfo{author}{Kim, D.~G.} \emph{et~al.}
\newblock \bibinfo{journal}{\bibinfo{title}{Social {Interaction} {Test} in
  {Home} {Cage} as a {Novel} and {Ethological} {Measure} of {Social} {Behavior}
  in {Mice}}}.
\newblock {\emph{\JournalTitle{Experimental Neurobiology}}}
  \textbf{\bibinfo{volume}{28}}, \bibinfo{pages}{247--260},
  \doiprefix\url{10.5607/en.2019.28.2.247} (\bibinfo{year}{2019}).

\bibitem{acikgoz_overview_2022}
\bibinfo{author}{Acikgoz, B.}, \bibinfo{author}{Dalkiran, B.} \&
  \bibinfo{author}{Dayi, A.}
\newblock \bibinfo{journal}{\bibinfo{title}{An overview of the currency and
  usefulness of behavioral tests used from past to present to assess anxiety,
  social behavior and depression in rats and mice}}.
\newblock {\emph{\JournalTitle{Behavioural Processes}}}
  \textbf{\bibinfo{volume}{200}}, \bibinfo{pages}{104670},
  \doiprefix\url{10.1016/j.beproc.2022.104670} (\bibinfo{year}{2022}).

\bibitem{puscian_blueprints_2022}
\bibinfo{author}{Puścian, A.} \& \bibinfo{author}{Knapska, E.}
\newblock \bibinfo{journal}{\bibinfo{title}{Blueprints for measuring natural
  behavior}}.
\newblock {\emph{\JournalTitle{iScience}}} \textbf{\bibinfo{volume}{25}},
  \doiprefix\url{10.1016/j.isci.2022.104635} (\bibinfo{year}{2022}).
\newblock \bibinfo{note}{Publisher: Elsevier}.

\bibitem{pearce_animal_2014}
\bibinfo{author}{Pearce, J.~M.}
\newblock \emph{\bibinfo{title}{Animal {Learning} and {Cognition}: {An}
  {Introduction}}} (\bibinfo{publisher}{Psychology Press},
  \bibinfo{address}{London}, \bibinfo{year}{2014}), \bibinfo{edition}{3} edn.

\bibitem{phifer-rixey_insights_2015}
\bibinfo{author}{Phifer-Rixey, M.} \& \bibinfo{author}{Nachman, M.~W.}
\newblock \bibinfo{journal}{\bibinfo{title}{Insights into mammalian biology
  from the wild house mouse {Mus} musculus}}.
\newblock {\emph{\JournalTitle{eLife}}} \textbf{\bibinfo{volume}{4}},
  \bibinfo{pages}{e05959}, \doiprefix\url{10.7554/eLife.05959}
  (\bibinfo{year}{2015}).
\newblock \bibinfo{note}{Publisher: eLife Sciences Publications, Ltd}.

\bibitem{bale_critical_2019}
\bibinfo{author}{Bale, T.~L.} \emph{et~al.}
\newblock \bibinfo{journal}{\bibinfo{title}{The critical importance of basic
  animal research for neuropsychiatric disorders}}.
\newblock {\emph{\JournalTitle{Neuropsychopharmacology}}}
  \textbf{\bibinfo{volume}{44}}, \bibinfo{pages}{1349--1353},
  \doiprefix\url{10.1038/s41386-019-0405-9} (\bibinfo{year}{2019}).
\newblock \bibinfo{note}{Number: 8 Publisher: Nature Publishing Group}.

\bibitem{homberg2013measuring}
\bibinfo{author}{Homberg, J.~R.}
\newblock \bibinfo{journal}{\bibinfo{title}{Measuring behaviour in rodents:
  towards translational neuropsychiatric research}}.
\newblock {\emph{\JournalTitle{Behavioural brain research}}}
  \textbf{\bibinfo{volume}{236}}, \bibinfo{pages}{295--306}
  (\bibinfo{year}{2013}).

\bibitem{peters_ethological_2015}
\bibinfo{author}{Peters, S.~M.}, \bibinfo{author}{Pothuizen, H. H.~J.} \&
  \bibinfo{author}{Spruijt, B.~M.}
\newblock \bibinfo{journal}{\bibinfo{title}{Ethological concepts enhance the
  translational value of animal models}}.
\newblock {\emph{\JournalTitle{European Journal of Pharmacology}}}
  \textbf{\bibinfo{volume}{759}}, \bibinfo{pages}{42--50},
  \doiprefix\url{10.1016/j.ejphar.2015.03.043} (\bibinfo{year}{2015}).

\bibitem{kondrakiewicz_ecological_2019}
\bibinfo{author}{Kondrakiewicz, K.}, \bibinfo{author}{Kostecki, M.},
  \bibinfo{author}{Szadzińska, W.} \& \bibinfo{author}{Knapska, E.}
\newblock \bibinfo{journal}{\bibinfo{title}{Ecological validity of social
  interaction tests in rats and mice}}.
\newblock {\emph{\JournalTitle{Genes, Brain and Behavior}}}
  \textbf{\bibinfo{volume}{18}}, \bibinfo{pages}{e12525},
  \doiprefix\url{10.1111/gbb.12525} (\bibinfo{year}{2019}).
\newblock \bibinfo{note}{\_eprint:
  https://onlinelibrary.wiley.com/doi/pdf/10.1111/gbb.12525}.

\bibitem{shemesh_paradigm_2023}
\bibinfo{author}{Shemesh, Y.} \& \bibinfo{author}{Chen, A.}
\newblock \bibinfo{journal}{\bibinfo{title}{A paradigm shift in translational
  psychiatry through rodent neuroethology}}.
\newblock {\emph{\JournalTitle{Molecular Psychiatry}}} \bibinfo{pages}{1--11},
  \doiprefix\url{10.1038/s41380-022-01913-z} (\bibinfo{year}{2023}).
\newblock \bibinfo{note}{Publisher: Nature Publishing Group}.

\bibitem{grieco_measuring_2021}
\bibinfo{author}{Grieco, F.} \emph{et~al.}
\newblock \bibinfo{journal}{\bibinfo{title}{Measuring {Behavior} in the {Home}
  {Cage}: {Study} {Design}, {Applications}, {Challenges}, and {Perspectives}}}.
\newblock {\emph{\JournalTitle{Frontiers in Behavioral Neuroscience}}}
  \textbf{\bibinfo{volume}{15}}, \bibinfo{pages}{735387},
  \doiprefix\url{10.3389/fnbeh.2021.735387} (\bibinfo{year}{2021}).

\bibitem{gunaydin2014natural}
\bibinfo{author}{Gunaydin, L.~A.} \emph{et~al.}
\newblock \bibinfo{journal}{\bibinfo{title}{Natural neural projection dynamics
  underlying social behavior}}.
\newblock {\emph{\JournalTitle{Cell}}} \textbf{\bibinfo{volume}{157}},
  \bibinfo{pages}{1535--1551} (\bibinfo{year}{2014}).

\bibitem{padilla2022dynamic}
\bibinfo{author}{Padilla-Coreano, N.}, \bibinfo{author}{Tye, K.~M.} \&
  \bibinfo{author}{Zelikowsky, M.}
\newblock \bibinfo{journal}{\bibinfo{title}{Dynamic influences on the neural
  encoding of social valence}}.
\newblock {\emph{\JournalTitle{Nature Reviews Neuroscience}}}
  \textbf{\bibinfo{volume}{23}}, \bibinfo{pages}{535--550}
  (\bibinfo{year}{2022}).

\bibitem{lee2021neural}
\bibinfo{author}{Lee, C.~R.}, \bibinfo{author}{Chen, A.} \&
  \bibinfo{author}{Tye, K.~M.}
\newblock \bibinfo{journal}{\bibinfo{title}{The neural circuitry of social
  homeostasis: Consequences of acute versus chronic social isolation}}.
\newblock {\emph{\JournalTitle{Cell}}} \textbf{\bibinfo{volume}{184}},
  \bibinfo{pages}{1500--1516} (\bibinfo{year}{2021}).

\bibitem{mathis_deeplabcut_2018}
\bibinfo{author}{Mathis, A.} \emph{et~al.}
\newblock \bibinfo{journal}{\bibinfo{title}{{DeepLabCut}: markerless pose
  estimation of user-defined body parts with deep learning}}.
\newblock {\emph{\JournalTitle{Nature Neuroscience}}}
  \textbf{\bibinfo{volume}{21}}, \bibinfo{pages}{1281--1289},
  \doiprefix\url{10.1038/s41593-018-0209-y} (\bibinfo{year}{2018}).
\newblock \bibinfo{note}{Number: 9 Publisher: Nature Publishing Group}.

\bibitem{graving_deepposekit_2019}
\bibinfo{author}{Graving, J.~M.} \emph{et~al.}
\newblock \bibinfo{journal}{\bibinfo{title}{{DeepPoseKit}, a software toolkit
  for fast and robust animal pose estimation using deep learning}}.
\newblock {\emph{\JournalTitle{eLife}}} \textbf{\bibinfo{volume}{8}},
  \bibinfo{pages}{e47994}, \doiprefix\url{10.7554/eLife.47994}
  (\bibinfo{year}{2019}).

\bibitem{pereira_fast_2019}
\bibinfo{author}{Pereira, T.~D.} \emph{et~al.}
\newblock \bibinfo{journal}{\bibinfo{title}{Fast animal pose estimation using
  deep neural networks}}.
\newblock {\emph{\JournalTitle{Nature Methods}}} \textbf{\bibinfo{volume}{16}},
  \bibinfo{pages}{117--125}, \doiprefix\url{10.1038/s41592-018-0234-5}
  (\bibinfo{year}{2019}).
\newblock \bibinfo{note}{Number: 1 Publisher: Nature Publishing Group}.

\bibitem{pereira_sleap_2022}
\bibinfo{author}{Pereira, T.~D.} \emph{et~al.}
\newblock \bibinfo{journal}{\bibinfo{title}{{SLEAP}: {A} deep learning system
  for multi-animal pose tracking}}.
\newblock {\emph{\JournalTitle{Nature Methods}}} \textbf{\bibinfo{volume}{19}},
  \bibinfo{pages}{486--495}, \doiprefix\url{10.1038/s41592-022-01426-1}
  (\bibinfo{year}{2022}).
\newblock \bibinfo{note}{Number: 4 Publisher: Nature Publishing Group}.

\bibitem{lehner_rats_2011}
\bibinfo{author}{Lehner, S.~R.}, \bibinfo{author}{Rutte, C.} \&
  \bibinfo{author}{Taborsky, M.}
\newblock \bibinfo{journal}{\bibinfo{title}{Rats {Benefit} from {Winner} and
  {Loser} {Effects}}}.
\newblock {\emph{\JournalTitle{Ethology}}} \textbf{\bibinfo{volume}{117}},
  \bibinfo{pages}{949--960}, \doiprefix\url{10.1111/j.1439-0310.2011.01962.x}
  (\bibinfo{year}{2011}).
\newblock \bibinfo{note}{\_eprint:
  https://onlinelibrary.wiley.com/doi/pdf/10.1111/j.1439-0310.2011.01962.x}.

\bibitem{pellis2022measuring}
\bibinfo{author}{Pellis, S.} \emph{et~al.}
\newblock \bibinfo{journal}{\bibinfo{title}{Measuring play fighting in rats: a
  multilayered approach}}.
\newblock {\emph{\JournalTitle{Current Protocols}}}
  \textbf{\bibinfo{volume}{2}}, \bibinfo{pages}{e337} (\bibinfo{year}{2022}).

\bibitem{ham2023goldilocks}
\bibinfo{author}{Ham, J.~R.} \& \bibinfo{author}{Pellis, S.~M.}
\newblock \bibinfo{journal}{\bibinfo{title}{The goldilocks principle: Balancing
  familiarity and novelty in the selection of play partners in groups of
  juvenile male rats}}.
\newblock {\emph{\JournalTitle{Animal Behavior and Cognition}}}
  \textbf{\bibinfo{volume}{10}}, \bibinfo{pages}{304--328}
  (\bibinfo{year}{2023}).

\bibitem{fulenwider2022manifestations}
\bibinfo{author}{Fulenwider, H.~D.}, \bibinfo{author}{Caruso, M.~A.} \&
  \bibinfo{author}{Ryabinin, A.~E.}
\newblock \bibinfo{journal}{\bibinfo{title}{Manifestations of domination:
  Assessments of social dominance in rodents}}.
\newblock {\emph{\JournalTitle{Genes, Brain and Behavior}}}
  \textbf{\bibinfo{volume}{21}}, \bibinfo{pages}{e12731}
  (\bibinfo{year}{2022}).

\bibitem{ramos1997multiple}
\bibinfo{author}{Ramos, A.}, \bibinfo{author}{Berton, O.},
  \bibinfo{author}{Morm{\`e}de, P.} \& \bibinfo{author}{Chaouloff, F.}
\newblock \bibinfo{journal}{\bibinfo{title}{A multiple-test study of
  anxiety-related behaviours in six inbred rat strains}}.
\newblock {\emph{\JournalTitle{Behavioural brain research}}}
  \textbf{\bibinfo{volume}{85}}, \bibinfo{pages}{57--69}
  (\bibinfo{year}{1997}).

\bibitem{neumann_elochoice_2019}
\bibinfo{author}{Neumann, C.}
\newblock \bibinfo{title}{{EloChoice}: {Preference} {Rating} for {Visual}
  {Stimuli} {Based} on {Elo} {Ratings}} (\bibinfo{year}{2019}).

\bibitem{neumann_elosteepness_2023}
\bibinfo{author}{Neumann, C.}
\newblock \bibinfo{title}{{EloSteepness}: {Bayesian} {Dominance} {Hierarchy}
  {Steepness} via {Elo} {Rating} and {David}'s {Scores}}
  (\bibinfo{year}{2023}).

\end{thebibliography}

\clearpage
\newpage


\section*{}
\setcounter{figure}{0} 
\renewcommand{\thefigure}{S\arabic{figure}} 

\setcounter{table}{0} 
\renewcommand{\thetable}{S\arabic{table}} 

\begin{table}
	{\Large
		{\textbf{Supplementary information:}\\[3pt]
			\textbf{Long-term tracking of social structure in groups of rats}} 
	}
\end{table}
\renewcommand{\thefigure}{S\arabic{figure}}
\setcounter{figure}{0}

\begin{figure}
	\centering
	\includegraphics[width=0.6\linewidth]{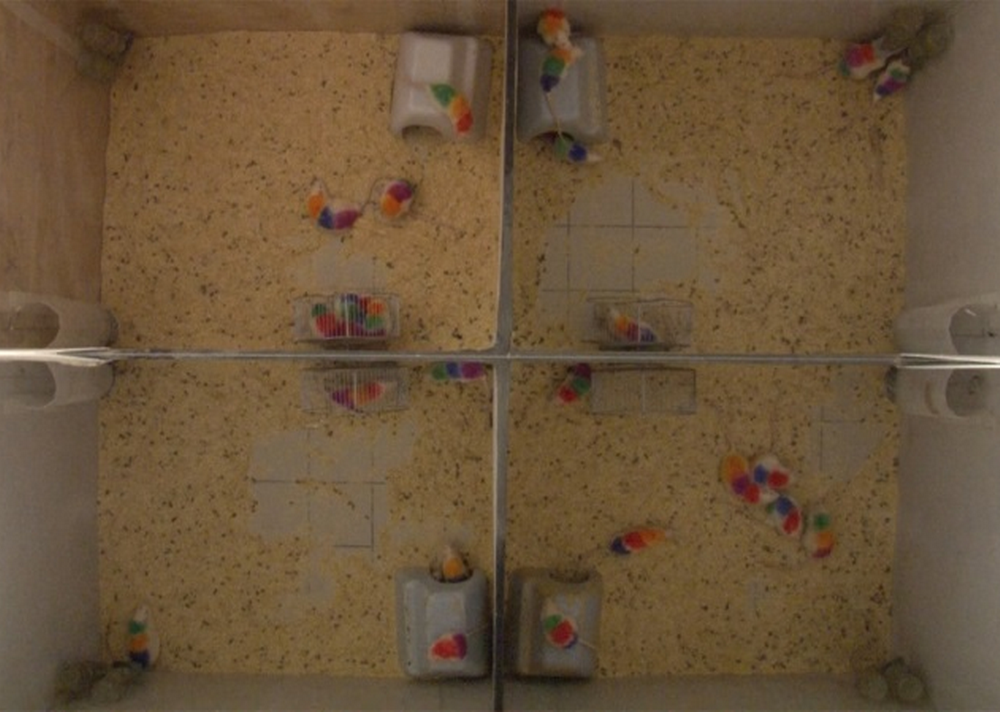}
	\caption{\textbf{Example camera frame image}.  Taken at low-light condition (i.e.\ during active period).}
	\label{sfig:exampleframe}
\end{figure}

\begin{figure}
	\centering
	\includegraphics[width=\linewidth]{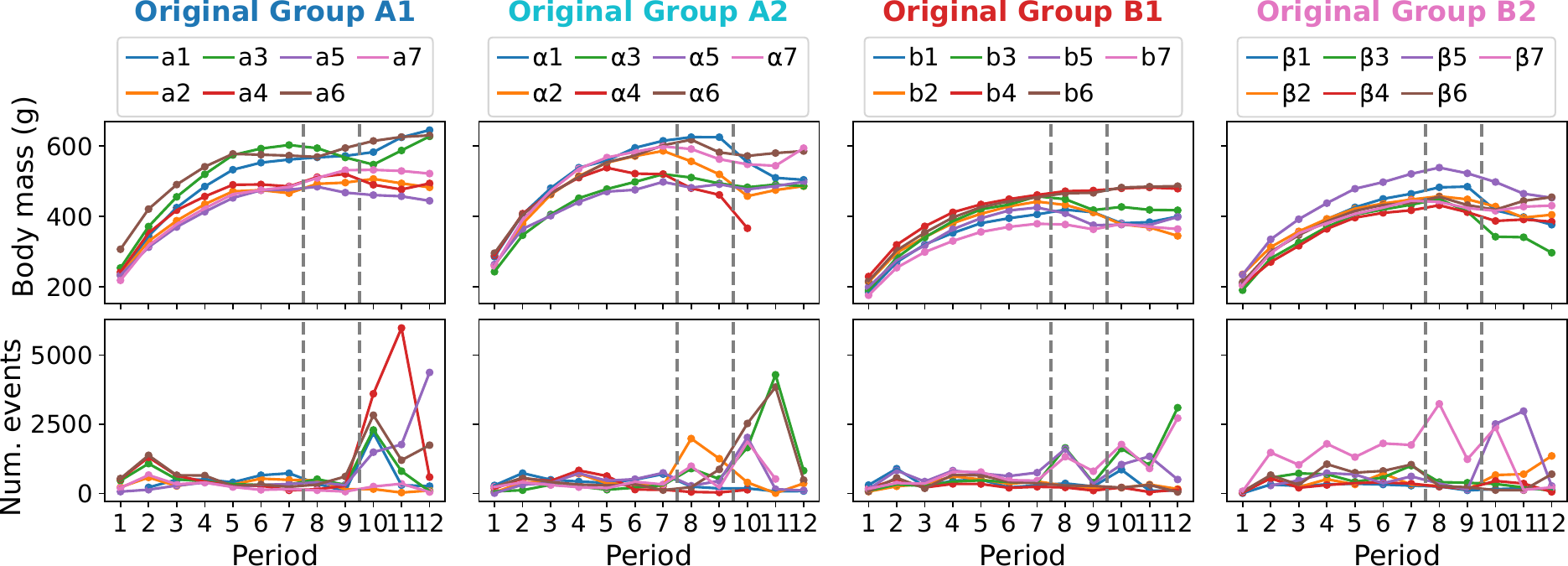}
	\caption{\textbf{Body mass and number of events during the entire experiment}.
		Individual rats are plotted according to phase 1 groups and shown with different colored lines.
		Num.\ events are the mean number of pairwise events one rat had with other rats in the same group.
	}
	\label{sfig:indivmetrics}
\end{figure}

\begin{figure}
	\centering
	\includegraphics[width=0.85\linewidth]{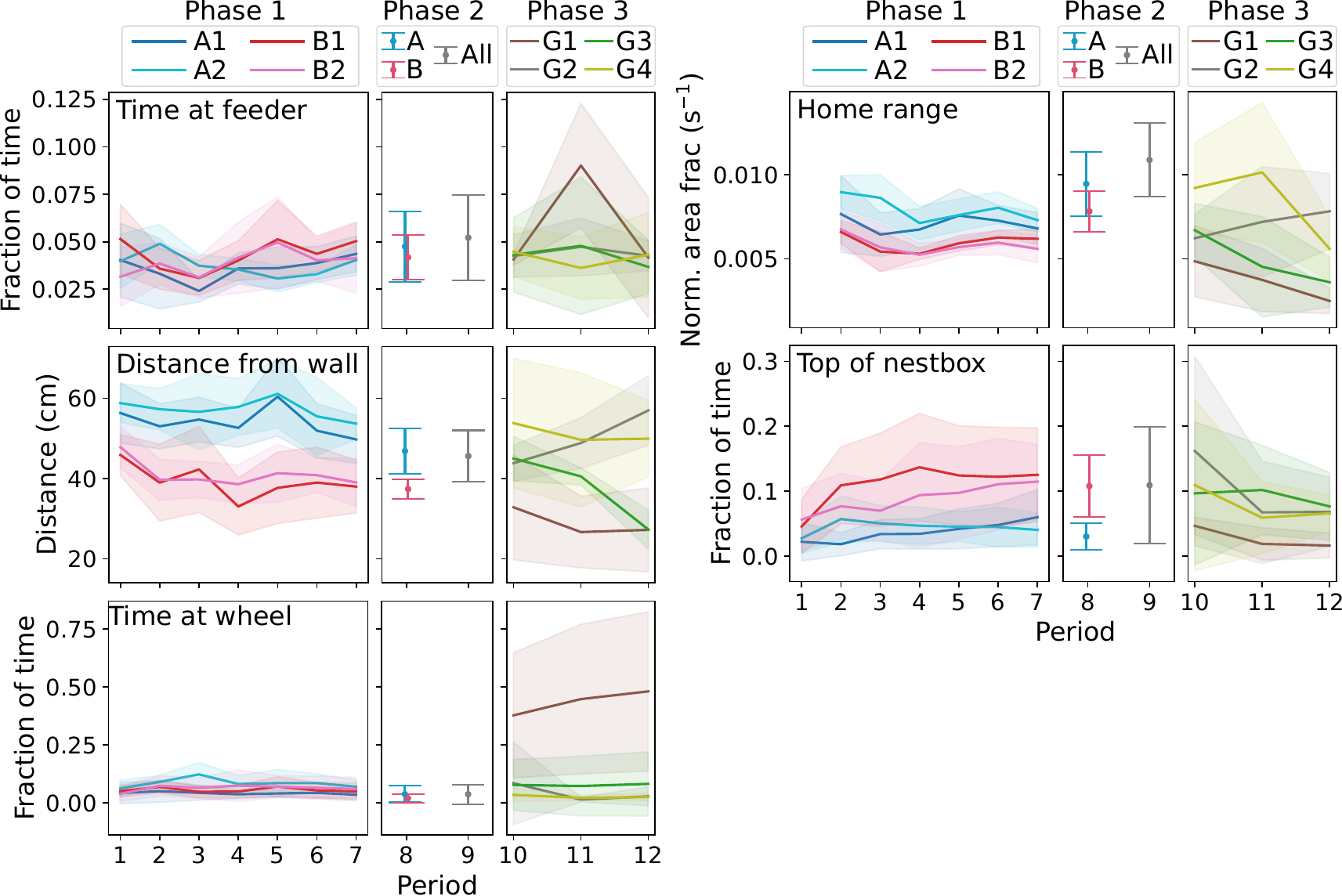}
	\caption{\textbf{Space use metrics for each group}.
		Metrics of time at feeder, distance from wall, home range, top of nestbox, and time at wheel.
		The lines/shaded area or points/error bars show the mean/standard deviation of each metric within each group, for the different phases.
		Note:\ home range was not calculated for Pd 1.   
	}
	\label{sfig:groupmetrics}
\end{figure}

\begin{figure}
	\centering
	\includegraphics[width=\linewidth]{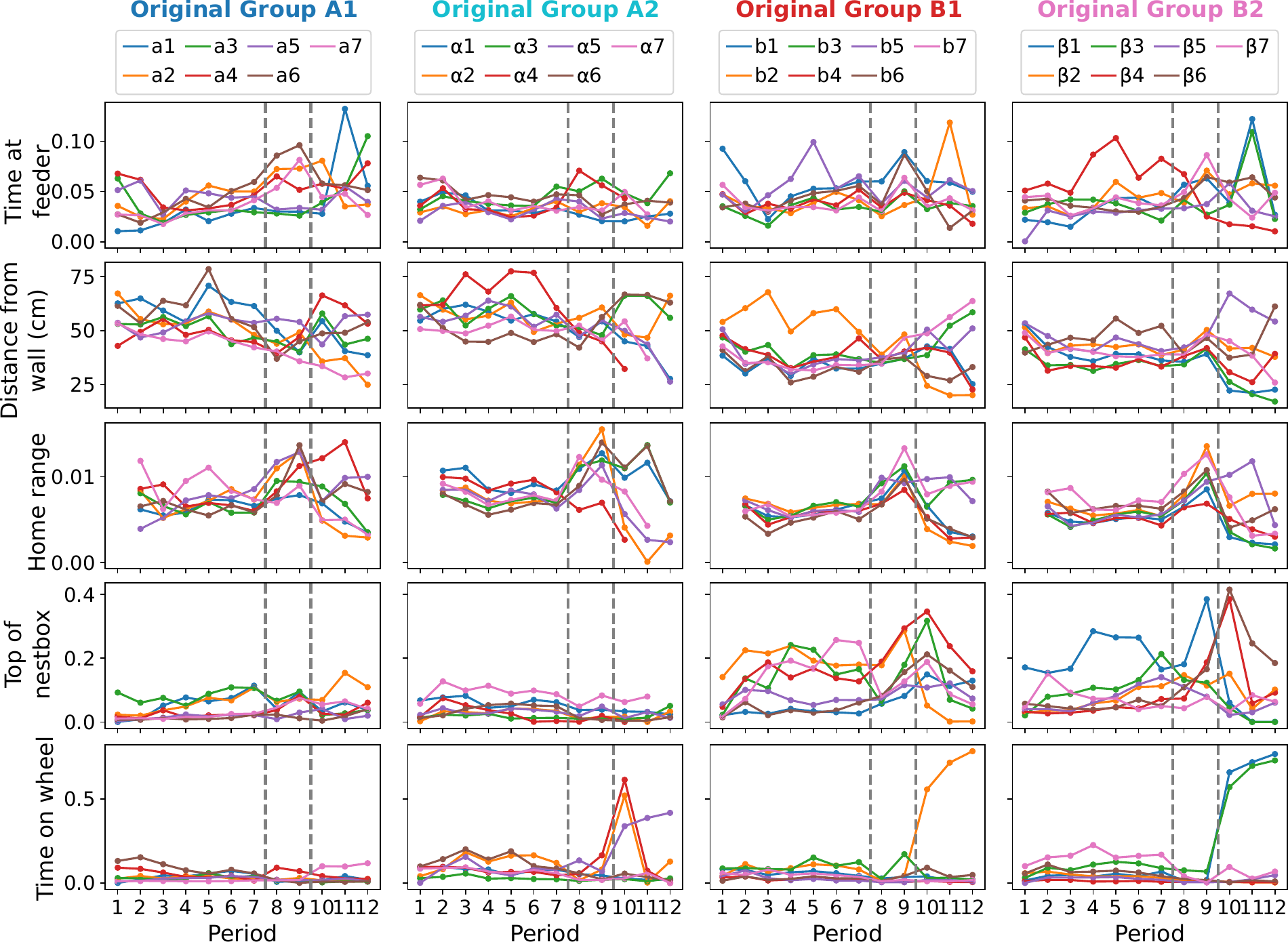}
	\caption{\textbf{Space use metrics plotted for individual rats}.
		See also Fig \ref{sfig:groupmetrics} for the average and standard deviation of space use metrics across individual rats in each group.
		Rats are plotted according to phase 1 groups in an analogous way to Fig \ref{sfig:indivmetrics}.
		Time at feeder is calculated as fraction of time, distance from wall has units of cm, home range is calculated as normalized area fraction (s$^{-1}$), and top of nestbox and time on wheel are are calculated as fraction of time.
	}
	\label{sfig:indivspaceuse}
\end{figure}

\begin{figure}
	\vspace{-45pt}
	\centering
	\includegraphics[width=0.8\linewidth]{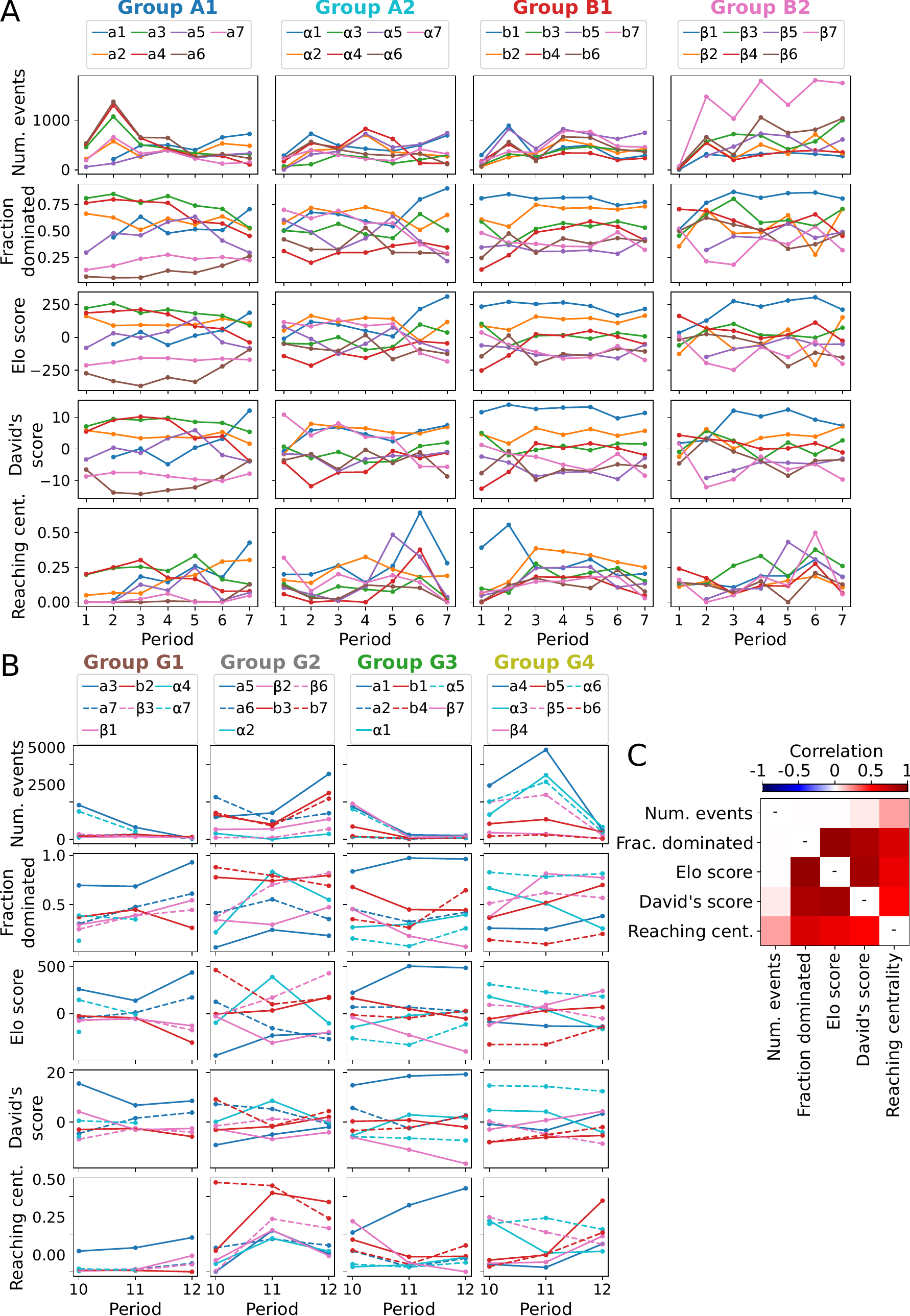}
	\caption{\textbf{Individual ranking metrics comparison and correlations.} 
		Shown are the total number of events for each rat, and metrics based on the matrix of pairwise event outcomes for rats in each group, including fraction of events dominated, Elo score, David's score, and the Local Reaching Centrality (LRC). 
		(A) Values for each rat according to phase 1 groups. 
		(B) Values for each rat according to phase 3 groups.  Colors designate which groups each rat belonged to during phase 1.
		(C) Correlation among metrics, calculated as the correlation coefficient for a given pair of metrics during the same period, using all data from phases 1 and 3.
	}
	\label{sfig:indivnetworkmetrics}
\end{figure}

\begin{figure}
	\centering
	\includegraphics[width=\linewidth]{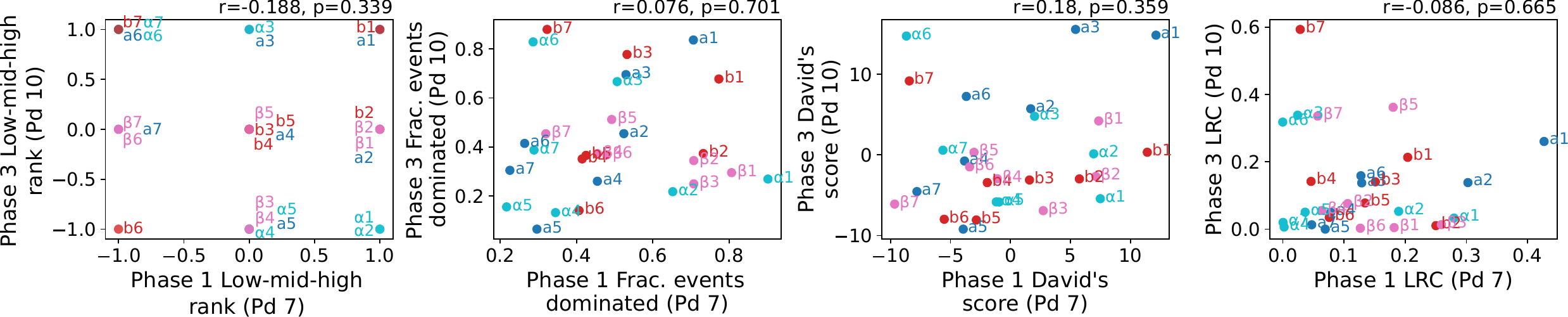}
	\caption{\textbf{Other phase 1 to phase 3 individual social ranking measures.}
		See Fig \ref{fig:bodymass} for phase 1 to phase 3 Elo scores for individual rats.  Shown here are ranks based on Elo scores as well as other individual rat ranking metrics, including fraction of events dominated, David's score, and local reaching centrality (LRC).
		Each plot compares scores at the end of phase 1 (Pd 7) to those at the start of phase 3 (Pd 10), i.e.\ after the new groups were formed.
		Low-mid-high ranks are determined by assigning a subordinate (low) value of -1 to the lowest two Elo scores in a group during a certain period, a dominant (high) value to the highest two Elo scores, and a middle (0) value to others.
	}
	\label{sfig:indivnetwork_p1_p3}
\end{figure}


\begin{figure}
	\centering
	\includegraphics[width=\linewidth]{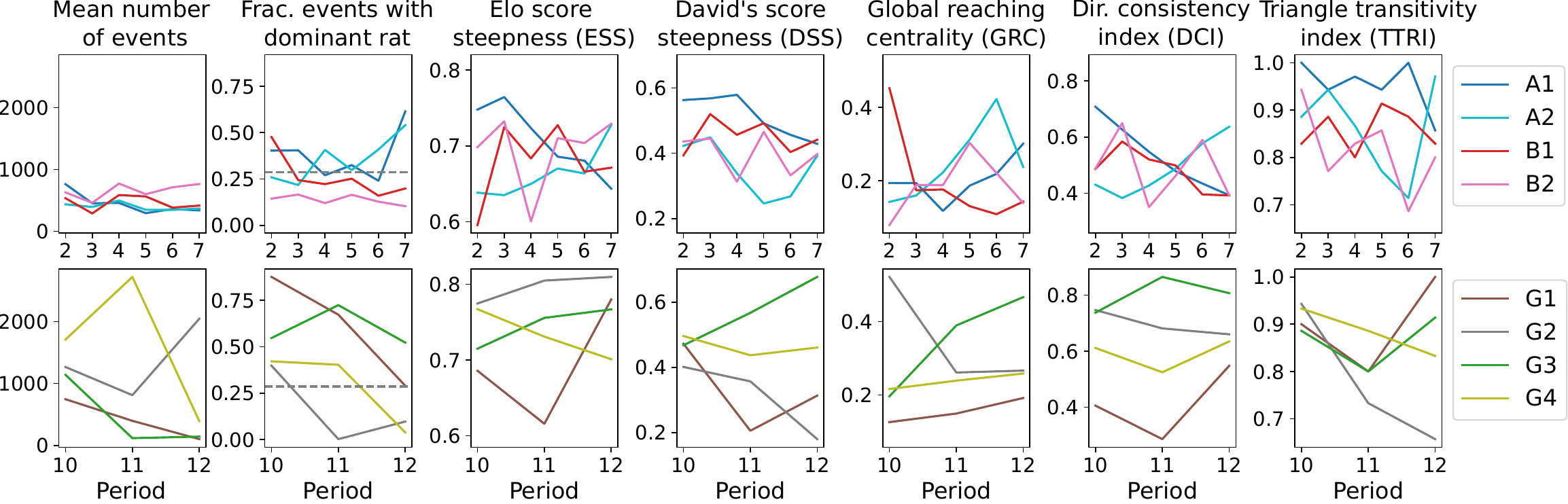}
	\caption{\textbf{Group social structure metrics over time}.
		See also Fig \ref{fig:groupnetwork} for boxplots of the same results, organized by group. The top row shows results for phase 1, and the bottom row for phase 3.
		The dashed line for fraction of events with dominant rat shows the expected value if all pairs of rats have the same number of events.
	}
	\label{sfig:gnetworkovertime}
\end{figure}

\begin{figure}
	\centering
	\includegraphics[width=\linewidth]{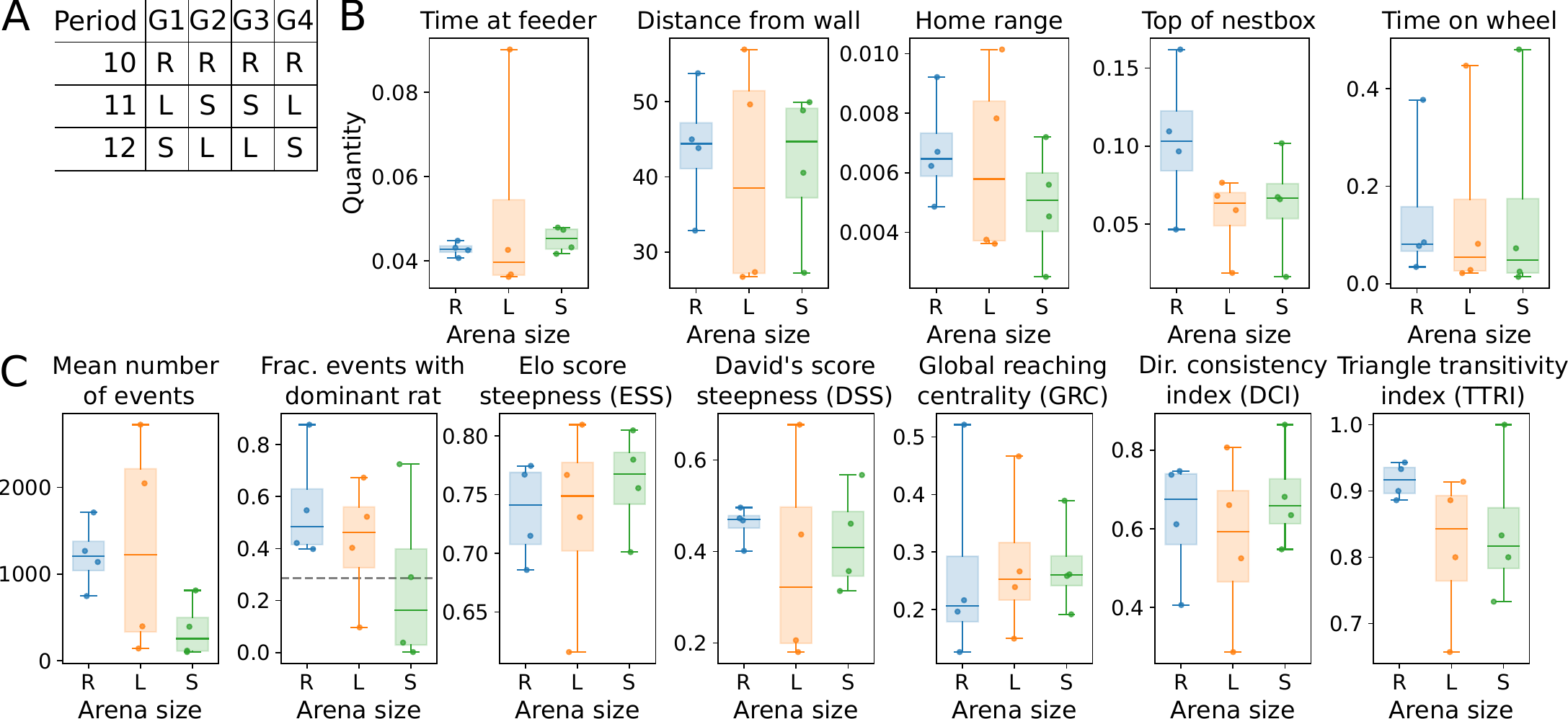}
	\caption{\textbf{Phase 3 living area size changes, space use, and group social structure metrics}.
		During phase 3, the size of the living area available to each group was changed by moving the barriers separating the compartments.  During Pd 10, all groups had the same Regular (R) - sized living area.  In Pds 11 and 12, the living areas were either large (L) or small (S).  
		(A) The living area sizes of each group during phase 3.  
		(B) Space-use metrics according to living area size.
		(C) Group social structure metrics according to living area size.
		The dashed line for fraction of events with dominant rat shows the expected value if all pairs of rats have the same number of events.    
	}
	\label{sfig:arenasize}
\end{figure}

\begin{figure}
	\centering
	\includegraphics[width=1\linewidth]{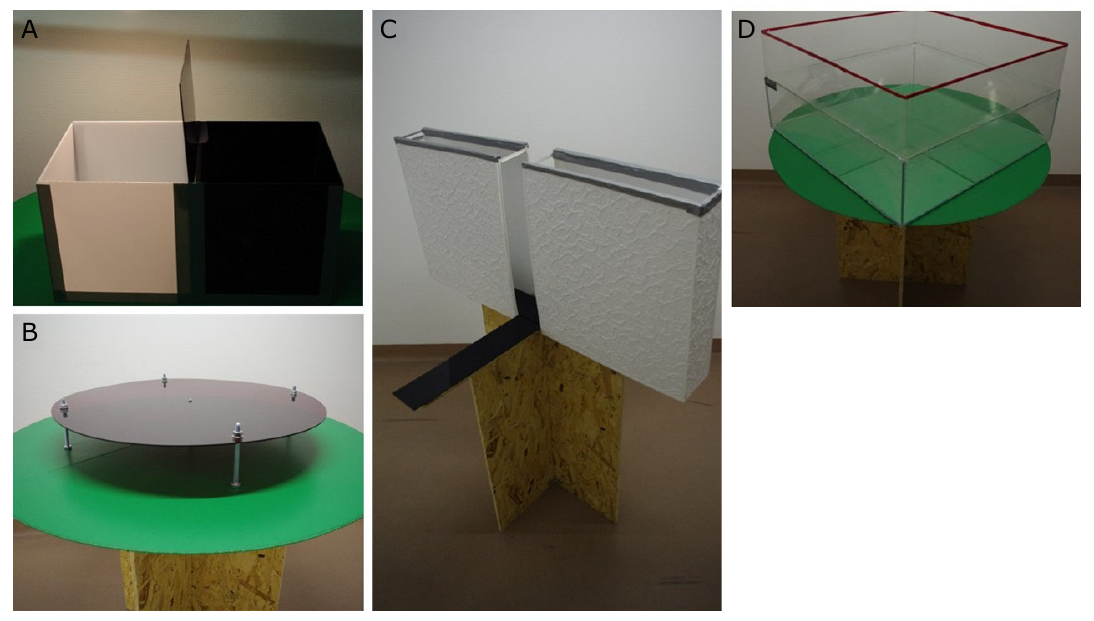}
	\caption{\textbf{Experimental setup for individual and social tests}. (A) black and white box, (B) canopy, (C) elevated plus-maze, and (D) test apparatus used for pairwise social tests.}
	\label{sfig:indiv_tests}
\end{figure}

\begin{figure}
	\centering
	\includegraphics[width=\linewidth]{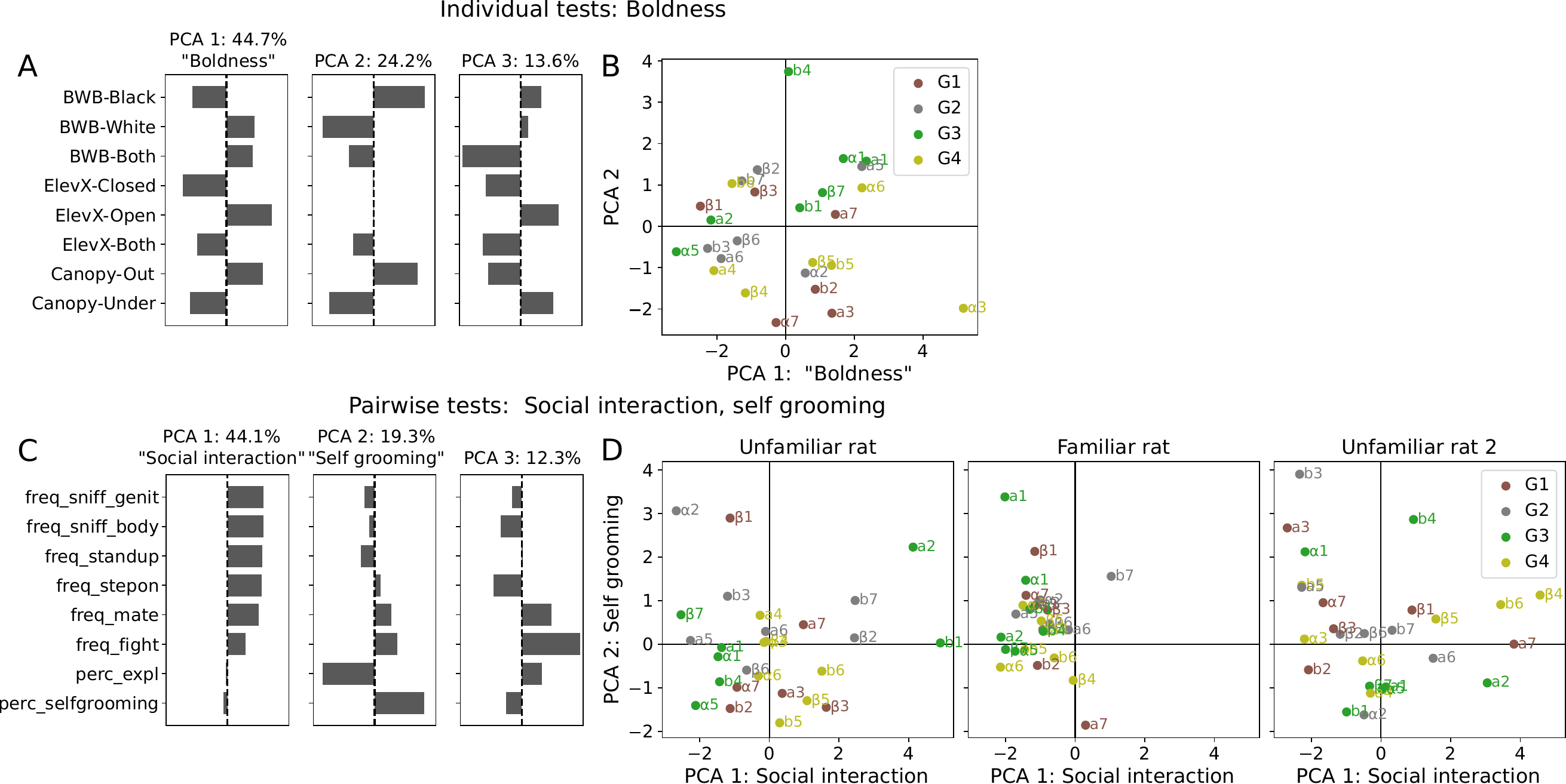}
	\caption{\textbf{PCA to get individual scores from assays.}
		(A) The black and white box, canopy, and elevated plus-maze test scores (Figure \ref{sfig:indiv_tests}A-C) were used as input to principal component analysis (PCA).  The first PCA component is used to define the composite ``Boldness'' score.
		(B) Embedding labeling each individual rat's score values projected onto the first two PCA axes, with colors representing the different phase 3 groups.
		(C) Measures from the pairwise interaction tests with an unfamiliar individual (Figure \ref{sfig:indiv_tests}D) were used as input to principal component analysis. Note that while the pairwise tests were also done with a familiar rat, and again with a different unfamiliar rat, the PCA components are determined and set from the first unfamiliar rat tests and these are shown here.  Projections onto the first PCA component are used as a ``Social interaction'' score, and onto the second component as a ``Self grooming'' score.
		(D) Embedding labeling each individual rat's scores projected onto the PCA axes shown in (C), for pairwise tests with an unfamiliar individual, a familiar individual, and a second unfamiliar individual. Colors represent the different phase 3 groups.
	}
	\label{sfig:assays_pca}
\end{figure}


\begin{figure}
	\centering
	\includegraphics[width=\linewidth]{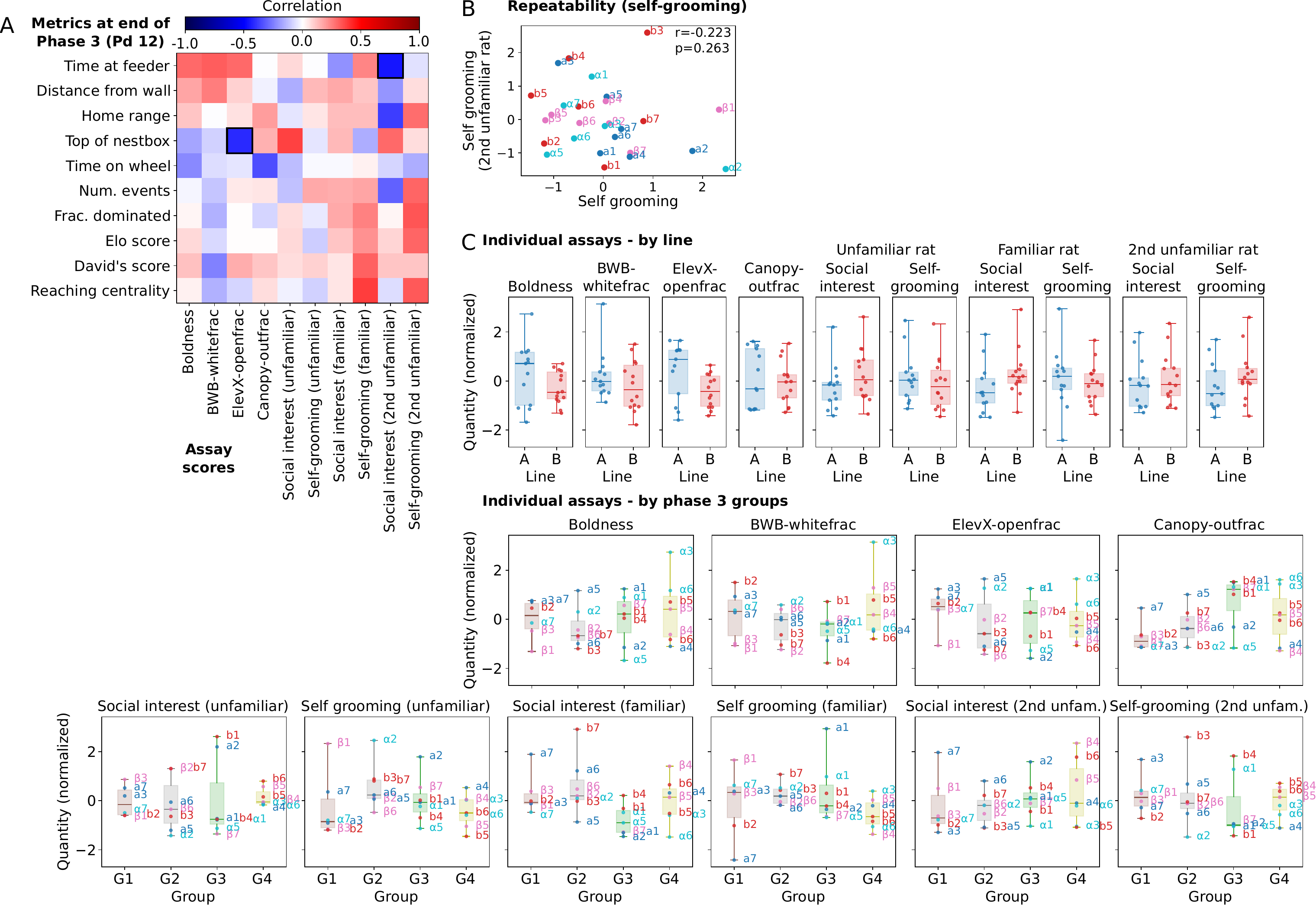}
	\caption{\textbf{Extended comparison of behavioral metrics to assays}.
		See Fig \ref{fig:indivtests} for focus comparison with Boldness and Social interaction scores.
		(A) Pearson correlation values for space use and social behavioral metrics from the final period in phase 3 (Pd 12) with individual assay scores.  Labels and color scales denotes correlation values. 
		Statistically significant correlations (p$>0.05$, calculated using t-distribution) are outlined in bold.
		(B) Comparison of self-grooming scores (pairwise tests, PCA 2 -- see Fig \ref{sfig:assays_pca}C) calculated from tests with a first unfamiliar rat (x-axis), with scores calculated from tests with a second unfamiliar rat (y-axis).  See  Fig \ref{fig:indivtests} for social interaction score (PCA 1).
		(C) Individual score distributions according to breeding line (left), and by phase 3 group membership (right). Scores are normalized by the mean and standard deviation of values measured for all rats.			
	}
	\label{sfig:indivtests_allscores}
\end{figure}

\begin{figure}
	\centering
	\includegraphics[width=\linewidth]{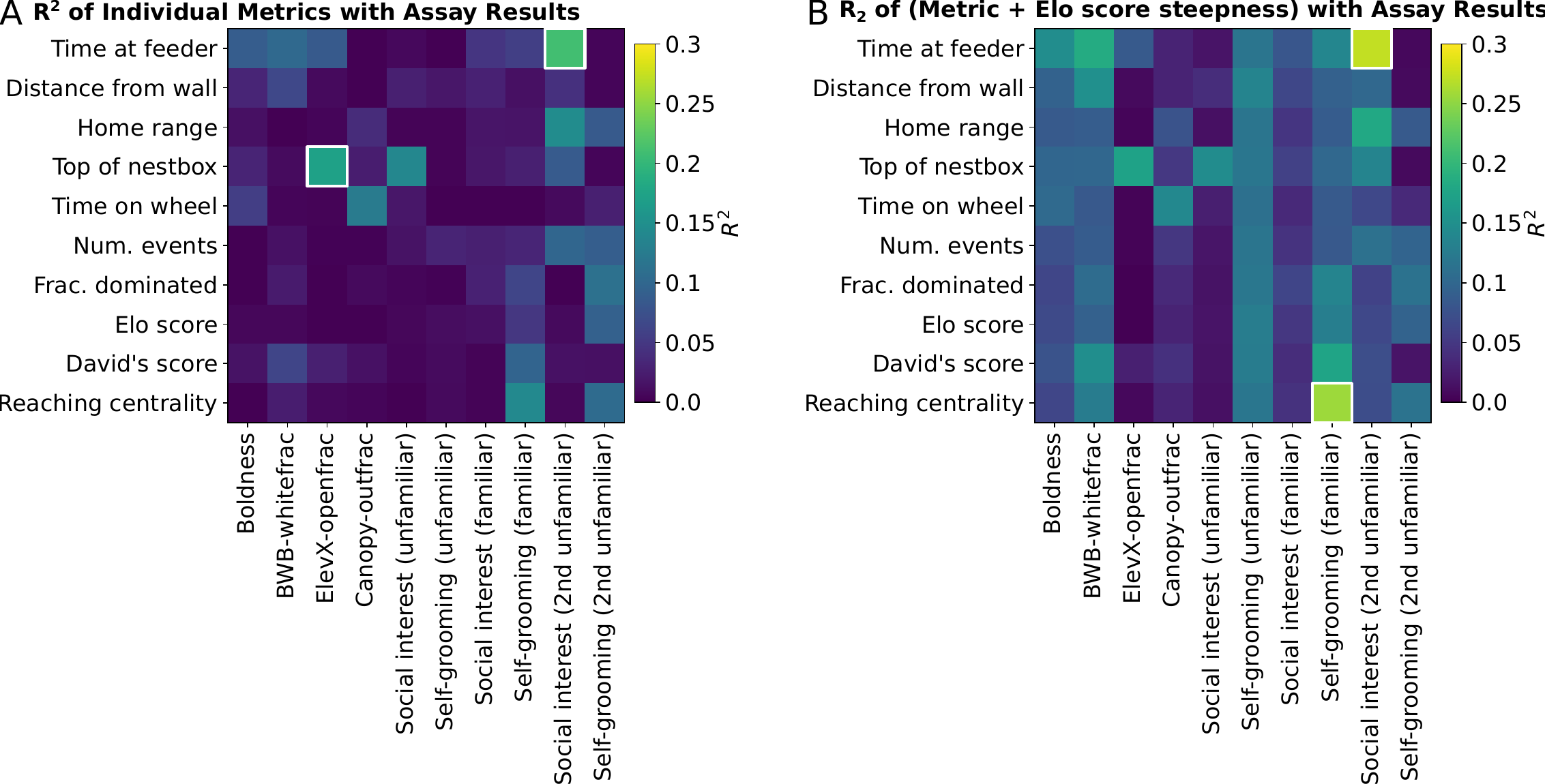}		
	\caption{\textbf{Interaction of individual and group social structure metrics in predicting individual assay results.}
		To assess whether interactions between individual and group social structure metrics enhance explanatory power in relation to the behavioral assays, we used linear regression to obtain $R^2$ values that quantify the proportion of variance explained by each model. Each plot shows $R^2$ values, with significant values (p$<$0.05, calculated using an F-test) outlined in white. The model fit results shown are:
		(A) Each metric fit separately: values displaying the explanatory power of individual space use and social behavioral metrics from phase 3 (Pd 12) for various assay scores.
		(B) A combined model with each metric and Elo score steepness as the two inputs.
	}
	\label{sfig:r2_metric_elo_steepness}
\end{figure}

\begin{figure}
	\centering
	\includegraphics[width=\linewidth]{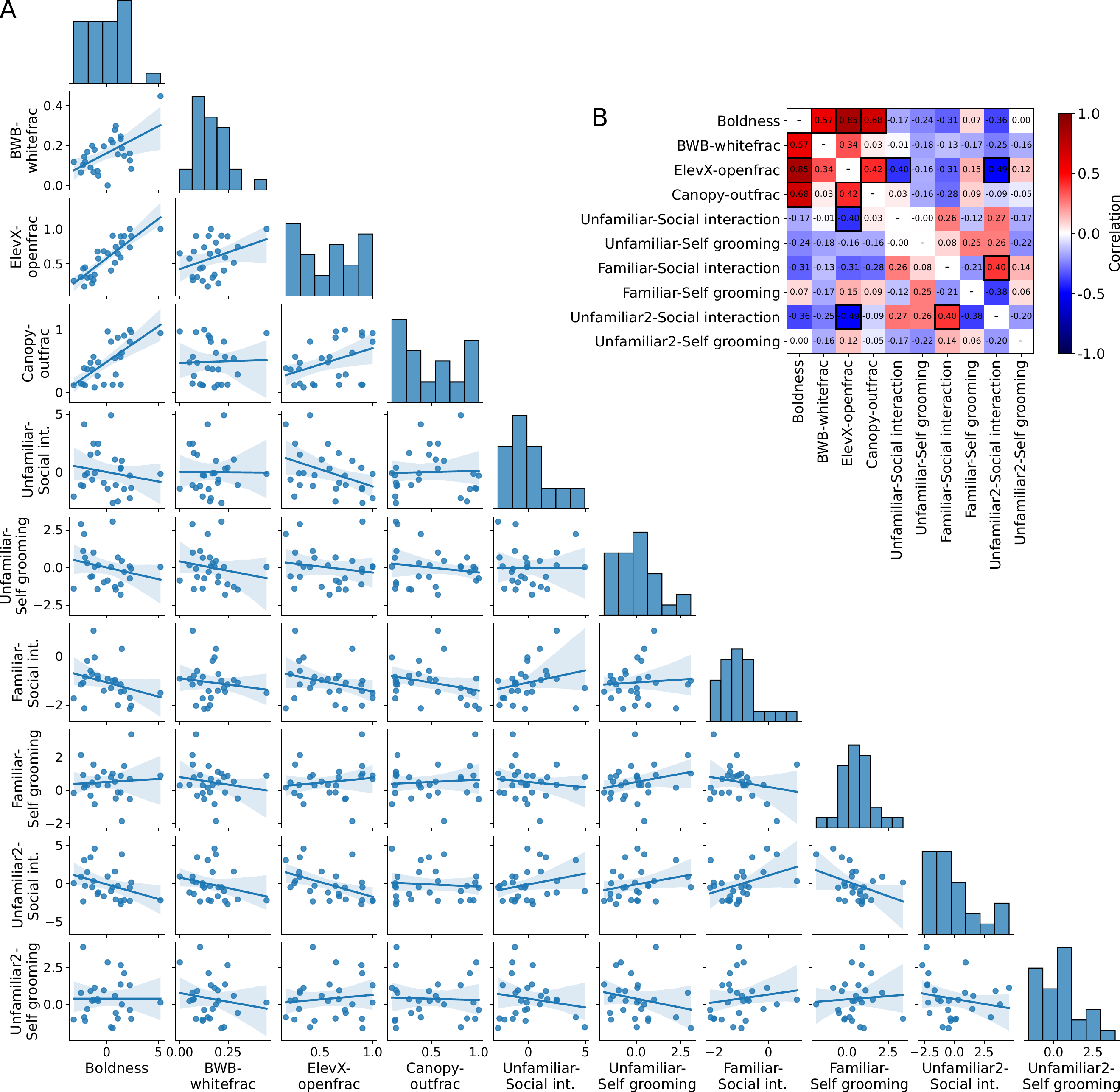}
	\caption{\textbf{Comparison of scores derived from individual and pairwise assays.}
		(A) Scatter plots with regression fit comparing scores derived from the individual and pairwise tests. The boldness score is defined using multiple tests (see Figure \ref{sfig:assays_pca}A), and here is also compared with results from each of these tests.  The social interaction and self grooming scores are defined using the pairwise tests (see Figure \ref{sfig:assays_pca}B).
		(B) Pearson correlation values of scores. The labels and colors indicate correlation values, and bold outlines highlight significant correlations ($p<0.05$, determined using t-distribution). 
	}
	\label{sfig:assay_score_comparison}
\end{figure}

\begin{figure}
	\caption{\textbf{Supplementary video:  Time-lapse video of selected days.}
		Video shows a time-lapse of recorded videos of rats in the experimental arena across 5 experimental days.  \href{https://hal.elte.hu/~nagymate/sjt/rat_social_data/}{Preview video online}.
	}
\end{figure}

\end{document}